\documentclass{emulateapj}

\usepackage{graphicx}

\shorttitle{Acceleration by MHD Turbulence}
\shortauthors{Cho \& Lazarian}

\begin{document}

\title{Particle Acceleration by MHD Turbulence}

\author{Jungyeon Cho}
\affil{Dept. of Astronomy and Space Science, Chungnam National University,
       Daejeon, Korea; cho@canopus.cnu.ac.kr}

\and 

\author{A. Lazarian}
\affil{Astronomy Dept., Univ.~of Wisconsin, Madison, 
       WI53706, USA; lazarian@astro.wisc.edu}

\begin{abstract}
Recent advances in understanding of magnetohydrodynamic (MHD)
turbulence call for revisions in the picture of particle acceleration.
We make use of the recently established scaling of slow and fast
MHD modes in strong and weak MHD turbulence to provide a systematic
study of particle acceleration 
in magnetic pressure (low-$\beta$) and gaseous pressure
(high-$\beta$) dominated plasmas. We consider the acceleration by large
scale compressions in both slow and fast particle
diffusion limits. We compare the results with the acceleration rate that
arises from resonance scattering and Transit-Time Damping (TTD).
 We establish that fast modes  
accelerate particles more efficiently than slow modes.
We find that particle acceleration by pitch-angle scattering and TTD 
dominates acceleration by
slow or fast modes when the spatial diffusion rate is small.
When the rate of spatial diffusion of particles is high, we
 establish an enhancement of the efficiency of
particle acceleration by slow and fast modes in weak turbulence.
We show that highly supersonic turbulence is an efficient
agent for particle acceleration.
We find that even incompressible turbulence can accelerate particles
on the scales comparable with the particle mean free path. 

\end{abstract}

                             

\section{\label{sec:intro}Introduction}

MHD turbulence is an important agent for particle acceleration
as was pointed first by Fermi (1949) and later was discussed by
many other authors (see Chandran \& Maron 2004b and references therein). 
Second order Fermi acceleration by MHD turbulence
was appealed for  acceleration of particles in many astrophysical
environments, e.g. Solar wind,  Solar 
flares, the intracluster medium, gamma-ray bursts (see Schlickeiser
\& Miller 1998; Chandran 2003; Petrosian \& Liu 2004).
Naturally, properties of MHD turbulence (see 
Shebalin, Matthaeus, \& Montgomery 1983; 
Higdon 1984; Montgomery, Brown, \& Matthaeus 1987;
Shebalin \& Montgomery 1988; Zank \& Matthaeus 1992; 
Cho \& Lazarian 2005 and references
therein) are essential for 
understanding the acceleration mechanisms.

For most part of the paper we use the model of {\it strong MHD}
turbulence initiated by
a pioneering work by 
Goldreich \& Sridhar (1995, henceforth GS95). GS95 dealt with
incompressible MHD turbulence and showed
Alfven  and pseudo-Alfven modes
follow the scale-dependent anisotropy of 
$l_{\|} \sim L^{1/3}l_{\perp}^{2/3}$,
where $l_{\|}$ is the size of the eddy along the local mean magnetic
field, $l_{\perp}$ that of the eddy perpendicular to it, and
$L$ the outer scale of turbulence.
Lithwick \& Goldreich (2001) conjectured that this scaling of incompressible
modes is also true for Alfven modes and slow modes in the
presence of compressibility. 
In Cho \& Lazarian (2002; 2003, henceforth CL03) we provided arguments justifying the rational
for considering separately the evolution of
slow, fast and Alfven mode cascades. 
Our numerical simulations\footnote{Higher
resolution numerics in Cho \& Lazarian (2004) confirms our earlier
results.} verified
that Alfven and slow mode velocity fluctuations indeed show 
GS95 scaling, while fast modes exhibit isotropy
in both gas-pressure (high $\beta$) and 
magnetic-pressure (low $\beta$) dominated plasmas (see review by
Cho \& Lazarian 2005).

MHD turbulence may also be in {\it weak} regime (Galtier et al. 2000). 
In weak MHD turbulence the
cascade increases only the perpendicular wavenumber and does not change
the parallel one. The inertial range of the weak turbulence in most 
astrophysical circumstances is limited. As the turbulence gets more and
more anisotropic it transfers into the strong MHD turbulence\footnote{The
names {\it strong} and {\it weak} may be misleading therefore. Strong
MHD turbulence may be of very small amplitude. Its characteristic property
is a so-called {\it critical balance}, which is the balance between the
time scales of perpendicular and parallel motions, i.e. 
$l_{\bot}/v_{\bot}\sim  l_{\|}/v_A$, where $v_A=B_0/\sqrt{4\pi\rho}$ is Alfven speed. Here $B_0$ is the strength of the mean magnetic field
and $\rho$ is density.
The weak turbulence decreases the perpendicular scale of the eddies
$l_{\bot}$ while keeping the parallel scale $l_{\|}$ intact. This
process can proceed till the critical balance condition is reached.
After that the turbulence is in the strong MHD regime
and both $l_{\bot}$ and $l_{\|}$ change in a way to preserve
the critical balance.  These issues are discussed,
for instance, in Cho, Lazarian \& Vishniac (2003).}. In spite of the 
limited inertial range, the weak turbulence may be important for cosmic
ray acceleration.

The new paradigm of MHD turbulence has substantially changed our
understanding of energetic particle-turbulence interactions via
gyroresonance and the  Transit-Time Damping (TTD) (Chandran 2000;
Yan \& Lazarian 2002, 2004; Farmer \& Goldreich 2004). However,
these two processes do not exhaust all the relevant interactions.
For instance, the acceleration of cosmic rays by 
the large scale compressible motions was described in the
literature rather long ago (see Ptuskin 1988, henceforth P88). 
Attempts of applying the GS95 scaling
to the
stochastic large-scale MHD acceleration of cosmic rays 
were described in some recent publications (Chandran 2003, henceforth C03; 
Chandran \& Maron 2004ab). The results were applied to particle acceleration
in Solar Flares and clusters of galaxies.

These studies raise a number of questions. For instance, it was found in
Yan \& Lazarian (2002, henceforth YL02) that fast modes dominate 
scattering of cosmic rays
if turbulent energy is injected at a large scale. Do the slow modes,
nevertheless, dominate the particle acceleration in Solar flares? 
What is the relative contribution of large scale compressions and small
scale scattering to the particle acceleration? Do the results change
if the turbulence is weak (see Galtier et al. 2000)? Is compressibility
a necessary requirement for acceleration? Is there a possibility to 
reconcile results by P88 and C03? These and other questions induce us
to undertake a detailed study of particle acceleration by MHD turbulence.

We consider particle acceleration by both strong and weak MHD turbulence.
We consider fast and slow modes, 
highly supersonic motions, effects of betatron electric field. We
 compare the rates of acceleration by different non-resonant
mechanisms and  the acceleration rates arising from pitch-angle 
scattering and TTD.

For the most part of the paper,
we consider a fluid threaded by a strong 
mean magnetic field ${\bf B}_0$. But the results may easily
be generalized if local magnetic field is used instead of
${\bf B}_0$ (see CL03).  

We start with introducing of the basics of the non-resonant 
acceleration by compressible turbulence in \S2. In the
same section we discuss particle acceleration in acoustic turbulence.
We consider particle acceleration by slow modes (\S3) and 
fast modes (\S4), 
highly supersonic turbulence (\S5), incompressible turbulence (\S6), and 
weak turbulence (\S7). We discuss gyroresonance and TTD acceleration 
in \S8 and compare the efficiency of different mechanisms in \S9.
We provide discussion in \S10 and summary in \S11.

\section{Acceleration by compressible motions: hydrodynamic approach}
\label{sect_hydro}

\subsection{Basics of non-resonant acceleration}   \label{sect_hydro_method}

The acceleration by large-scale motions 
depends on compression created by different modes. The 
traditional method of estimating $D_p$ is based on calculation
of the degree of fluid compression within a hydrodynamic approach.
It is implicitly assumed that the motions in question
are much larger than a particle mean
free path, the magnetic field moves together with the fluid and particles are
moved together with the magnetic field.

When a particle moves inside an eddy of size $l$, the
change of particle's momentum over a time $\Delta t$ is 
\begin{equation}
 \Delta p \sim (dp/dt)\Delta t,
\end{equation}
where $p$ is the momentum of the particle.
The momentum diffusion coefficient is
\begin{equation}
  D_p \sim  (\Delta p)^2/\Delta t \sim (dp/dt)^2 \Delta t.
  \label{eq_A2}
\end{equation}
The change of momentum depends on compression \footnote{
    We will discuss why 
    the change of momentum depends on $\nabla \cdot {\bf v}$ in \S\ref{sect_23}.
    We can understand it as follows.
    First, consider compression only in the directions perpendicular to
    the mean magnetic field. This will result in compression of magnetic field
    lines, which increases the perpendicular momentum of the particle.
    Second, consider compression only in the directions parallel to
    the mean magnetic field. This will cause adiabatic heating and 
    increase the parallel momentum of
    the particle.
    Overall, the rate of momentum change depends on the rate of compression.
}:
\begin{equation}
 \frac{ dp }{ dt } \approx -p \frac{ \nabla \cdot {\bf v}_l }{ c_0 },
  \label{eq_A3}
\end{equation}
where a constant $c_0$ depends on the geometry of compression.
For example, $c_0 = 3$ when compression is isotropic (see P88).
Here, for the sake of simplicity, we set $c_0=1$.
Substituting eq. (\ref{eq_A3}) into eq. (\ref{eq_A2}),
we obtain
\begin{equation}
  D_p \sim   p^2(\nabla \cdot {\bf v}_l)^2 \Delta t.
  \label{eq_univ}
\end{equation}
Therefore we can determine $D_p$ when we know
$\nabla \cdot {\bf v}_l$ and $\Delta t$.

\subsection{Wave-like turbulence}   \label{sect_32}
An important work for stochastic acceleration is P88,
that implicitly considers 
acoustic turbulence\footnote{Formally this corresponds to
weak isotropic turbulence caused by
interacting sound waves.
In this paper (see \S 7), {\it weak} turbulence
means that the wave-wave interaction 
is so weak that 
wave packets can travel distances much larger than their 
typical size.}.
Ptuskin compared two time scales on the outer scale
of turbulence - wave period ($t_{wave}\sim L/a$) 
and diffusion time ($t_{diff}\sim L^2/D$),
where $a$ is the sound speed, $D$ the 
spatial diffusion coefficient, and $L$ the outer scale of turbulence.

{\bf Fast diffusion limit}\\
When diffusion time, $t_{diff}$, is smaller than 
wave period, $t_{wave}$, we can use
$\Delta t =  t_{diff}$.
P88 considered isotropic turbulence arising from
interacting sound waves. For the outer scale eddies, 
since $\nabla \cdot {\bf v}_L \sim v_L/L$, we get
\begin{equation}
   D_p \sim p^2 (v_L^2/L^2)(L^2/D) \sim p^2 v_L^2/D,
   \label{eq_ptuskin_1app}
\end{equation}
where $v_L$ is the rms velocity
(P88).

It is straightforward to calculate $D_p$
for smaller eddies:
\begin{equation}
   D_p \sim \frac{ p^2 v_A }{ L } 
            \frac{ Lv_A }{ D }
            \left( \frac{ v_l }{ v_A } \right)^2    \label{eq_18l}
\end{equation}
for eddies on scale $l < L$, which is smaller than $D_p$ for outer scale 
eddies.
When we want to calculate contributions from all scale eddies, we need 
to calculate $\int D_p (dl/l)$:
\begin{equation}
  \int D_p (dl/l) \sim D_{p,L} \int (l/L)^{2m} (dl/l) 
                  \sim  D_{p,L}/2m,
\end{equation}
where $D_{p,L}$ is given in eq. (\ref{eq_ptuskin_1app}) and
we assumed $v_l \sim v_L (l/L)^m$.
Zakharov \& Sagdeev (1970) claimed 
that $m=1/4$, or
equivalently
$E(k)\propto k^{-3/2}$ for weak acoustic turbulence.
The result is very close to the one in eq. (\ref{eq_ptuskin_1app}).

{\bf Slow diffusion limit}\\
On the other hand, for $t_{diff} > t_{wave}$, 
P88 considered outer scale eddies and obtained
\begin{eqnarray}
    D_p  \sim p^2 (D v_L^2/a^2L^2)  \label{eq_ptuskin} \\
         \sim (p^2 v_L^2/a^2)/t_{L,diff} \label{eq_ptuskin_re} \\
         \sim  (p^2 v_L^2/D)(t_{L,wave}/t_{L,diff})^2
               \label{eq_ptuskin_re2}
\end{eqnarray}
which goes to zero when $D \rightarrow 0$.
This relation can be understood as follows.
Consider a particle inside a region of size $L$ that is compressed/expanded by large scale
wave packets 
in {\it weak} turbulence.
When the particle diffuses out of the region, the change of momentum
is $\Delta p \sim (dp/dt)t_{L,wave} \sim (p v_L/L)(L/a) \sim pv_L/a$.
However, since a particle
stays inside the region for approximately $t_{L,diff}$,
we take $\Delta t \sim t_{L,diff}$.
Substituting these $\Delta p$ and $\Delta t$ into 
$D_p\sim (\Delta p)^2/\Delta t$, we obtain
the result in eq. (\ref{eq_ptuskin_re}).

According to P88 original formula in eq. (\ref{eq_ptuskin}),
particle acceleration is inefficient in slow diffusion limit.
However, note that Ptuskin's formula in eq. (\ref{eq_ptuskin})
is derived for large scale eddies (or, to be precise, large scale compressive motions). 
It is easy to show that small scale eddies can provide efficient
acceleration. 
Let us consider eq. (\ref{eq_ptuskin_re2}).
For large scale eddies, the ratio $t_{L,wave}/t_{L,diff}$ is very small
in slow diffusion limit.
As we move down to smaller scale eddies, $t_{l,wave}$ scales as $l/a$ and
$t_{l,diff}$ as $l^2/D$. 
The ratio $t_{l,wave}/t_{l,diff}$ becomes larger as the scale
decreases. 
Therefore, as we move down to smaller scales,
we will have higher acceleration efficiencies.
When the inertial range is wide enough, we will ultimately reach
the scale $l_c$ on which $t_{l,wave} \sim t_{l,diff}$.
On the scale $l_c$, the momentum diffusion coefficient is
\begin{equation}
D_p\sim p^2v_l^2/D,
\end{equation}
where $v_l$ is the velocity at scale $l_c$

%
%

%

\section{Acceleration by compressible motions: slow modes} 
\label{sect_slow}

Here we consider acceleration by slow modes in moderately compressible 
turbulence for low Mach numbers. As above, we discuss both slow and fast
diffusion limits.
We assume turbulence is strong on the outer scale (i.e.~$v_L \sim v_A$).
In this case, the outer scale eddies are isotropic.
However, smaller eddies are elongated along the local mean magnetic field
and we have $l_{\|} \sim L^{1/3}l_{\perp}^{2/3}$, which
states that anisotropic eddy shape is more pronounced 
on smaller scales.
We consider separately gas pressure (high-$\beta$) and magnetic
pressure (low-$\beta$) dominated plasmas.
As in C03, we assume that spatial diffusion of particle occurs
due to pitch-angle scattering.

\subsection{Fast diffusion limit in low-$\beta$}    \label{sect_3_1}

Eq. (\ref{eq_univ}) is a universal formula for $D_p$ and
we can use it for slow modes.
Evaluation of eq. (\ref{eq_univ}) requires
$\nabla \cdot {\bf v}_{l,slow}$ and $\Delta t$.
Since we consider the case of $t_{L,diff} < t_{L, eddy}$, 
where $t_{L,eddy}\sim L/v_A$ is
the eddy turnover time on the outer scale,
particles diffuse out of an eddy
before it is randomized. 
Therefore, we take the diffusion time for $\Delta t$: 
$\Delta t \sim L^2/D_{\|}$, where we used $D_{\|}$ because
particle diffusion occurs along the magnetic field lines.
For slow modes in low-$\beta$ limit, we have
$\nabla \cdot {\bf v}_{L,slow} \sim v_{L,slow}/L_{\|}$ (see Appendix B).
When turbulence is isotropic on the outer scale, 
as assumed in this paper, we have $L_{\|} \sim L$.
The momentum diffusion coefficient is
\begin{eqnarray}
   D_p^{slow} \sim p^2 (v_{L,slow}/L)^2(L^2/D_{\|}) \nonumber \\
       \sim \frac{ p^2 v_A }{ L } 
            \frac{ Lv_A }{ D_{\|} }
            \left( \frac{ v_{L,slow} }{ v_A } \right)^2, 
       \label{eq_sm_fd1}
\end{eqnarray}
which is very close to eq. (\ref{eq_ptuskin_1app}) for
acoustic turbulence.
It is also straightforward to calculate $D_p$
for smaller eddies:
\begin{equation}
   D_{p,l} \sim \frac{ p^2 v_A }{ L } 
            \frac{ Lv_A }{ D_{\|} }
            \left( \frac{ v_{l,slow} }{ v_A } \right)^2    
    \label{eq_sm_fd_l}
\end{equation}
for eddies on scale $l < L$, which is smaller than $D_p$ for outer scale 
eddies. All these results are similar to those for acoustic
turbulence.

The only difference stems from the fact that when particles diffuse along
magnetic field lines, they can reenter the same outer scale eddy (Chandran
\& Maron 2004b). This results in a somewhat uncertain factor $N$ which
ranges from unity to several (see discussion in the Appendix A). 
This results in
\begin{equation}
  D_p \sim \frac{ p^2 v_A }{ L } 
            \frac{ Lv_A }{ D_{\|} }
            \left( \frac{ v_{l,slow} }{ v_A } \right)^2 N
\label{27}
\end{equation}
for eddies on scale $l$.

\subsection{Slow diffusion limit in low-$\beta$}
Consider $d_{min}^{}v_A \ll D_{\|}\ll L v_A$, 
where
$d_{min}$ is either the parallel size of eddies at the dissipation scale
or the mean free path of the cosmic rays, 
whichever is larger.
The particle diffusion time is smaller than the eddy turnover time at the 
outer scale $L$ and larger than that on the scale $\sim d_{min}$.
When $D_{\|} \ll Lv_A$, particle diffusion time $t_{L,diff}$ 
is larger than the eddy turnover time $t_{L,eddy}$ at the 
outer scale $L$.
In this case, 
for eddies whose parallel size is larger than $D_{\|}/v_A$, 
particles are confined within the eddies until the eddies are
randomized.

For this regime 
C03 obtained (see Appendix A)
\begin{equation}
   D_{p,l}^{} \sim \frac{ p^2 v_A }{ L }  \left(\frac{v_{L,slow}}{v_A}\right)^2
\label{eq_main_l}
\end{equation}
for a scale $l$ and
\begin{equation}
   D_p^{} \sim \frac{ p^2 v_A }{ L }  
        \left(\frac{v_{L,slow}}{v_A}\right)^2
    \ln\left( \frac{ Lv_A }{ D_{\|} } \right)
\label{eq_main_tot}
\end{equation}
for action of all eddies.
We observe, that 
each eddy whose parallel size is between $D_{\|}/v_A$ and $L$ 
makes an equal contribution to $D_p$. More importantly, the
efficiency of acceleration becomes large as $D_{\|} \rightarrow 0$.

On the contrary, when we apply the approach 
in \S\ref{sect_32} (see discussion
below eq. (\ref{eq_ptuskin_re2}))
to slow modes, we obtain\footnote{When the particle diffuses out of the eddy, the change of momentum
is $\Delta p \sim (dp/dt)t_{l,eddy} \sim (pv_{l,slow}/l_{\|})
    (l_{\|}/v_A)$.
The time that the particle spends
inside the eddy is $\sim t_{l,diff}$, so that we take
$\Delta t \sim t_{l,diff}$.
Substituting these $\Delta p$ and $\Delta t$ into 
$D_p\sim (\Delta p)^2/\Delta t$, we obtain
$D_p \sim (p^2v_{l,slow}^2/v_A^2)(D_{\|}/l_{\|}^2)
     \sim p^2 (D_{\|}v_{l,slow}^2/v_A^2 l_{\|}^2)$,
which is, in many aspects, similar to eq. (\ref{eq_ptuskin}).}
$D_p \sim p^2 (D_{\|}v_{l,slow}^2/v_A^2 l_{\|}^2)$.
As in acoustic turbulence, we have $D_p \rightarrow 0$
as $D_{\|} \rightarrow 0$. 
Note that the apparent disagreement between P88 and C03 does
not stem from isotropy/anisotropy of turbulence.

C03 assumes that the
random walk argument is applicable to the momentum
diffusion. This is a somewhat subtle issue. 
For instance, this is not true when diffusion time is much longer than
the eddy turnover time.
Let us consider an extreme case of $D_{\|} \rightarrow 0$.
In this case, the particle and the fluid element move together.
The density of the fluid element follows the continuity equation:
\begin{equation}
    \frac{ \partial\rho }{ \partial t } 
           = -{\bf v}\cdot \nabla \rho - \rho \nabla \cdot
    {\bf v}.
\end{equation}
Rearranging this, we get
\begin{equation}
    \frac{ 1 }{ \rho }\frac{ D\rho }{ Dt } = \frac{ D\ln \rho }{ Dt }
    = - \nabla \cdot {\bf v},
\end{equation}
where $D/Dt = \partial/\partial t + {\bf v}\cdot \nabla$.
We use $D/Dt$ because particle is moving with the fluid in the limit of
$D_{\|} \rightarrow 0$.
Since
\begin{equation}
     - \nabla \cdot {\bf v} = 
   \frac{ 1 }{ p }\frac{ Dp }{ Dt } = \frac{ D\ln p }{ Dt },
\end{equation}
where $p$ is the momentum of the particle,
$\ln p$ and $\ln \rho$ behave similarly.
Since $\ln \rho$ cannot go to infinity,
$\ln p$ cannot increase indefinitely.
See Appendix B for more discussion about the density fluctuations 
in compressible fluids.
Therefore, random walk may not be a good approximation for 
slow diffusion limit. This is a kind of suppression of random walk
in extremely slow diffusion limit.

{\bf Suggested Solution}---The apparent disagreement of earlier works
may be reconciled when we consider turbulent diffusion.
Here we mostly talk about diffusion in the directions perpendicular to
the mean field.
In weak turbulence turbulent diffusion is negligible and mobility
of particles relative to the rest of the fluid 
relies on $D_{\|}$ only.
Therefore we can apply our argument in the previous subsection and
$D_p$ goes to zero as $D_{\|}$ goes to zero.
However, in strong MHD turbulence, turbulent diffusion can be efficient
and a particle can move to another eddy within one eddy turnover time.
Recent numerical simulations (Cho et al.~2003) support the idea that turbulent 
diffusion is indeed efficient. However, diffusion in
MHD turbulence is a complicated issue that critically depends on
mobility of magnetic field lines and, hence, 
the efficiency of magnetic reconnection in turbulent environments
(see Lazarian \& Vishniac 1999). 
Therefore, we will not discuss this further here.

If turbulent diffusion is fast, particles can move to another
independent eddy within one eddy turnover time and random walk-like 
behavior will be fully recovered. The result will be the same as
the one in eq. (\ref{eq_main_tot}).

If turbulent diffusion is slow 
we can still have an efficient
acceleration 
(see a similar argument for hydrodynamic wave-like turbulence in \S\ref{sect_32}).
Indeed, for slow turbulent diffusion,
the eddies on the energy injection scale
induce  smaller $D_p$ than the estimate in eq. (\ref{eq_main_l}).
We expect that the suppression of random walk will be less severe
for smaller scale eddies because the inequality, $D_{\|} \ll l_{\|}v_A$
becomes milder for smaller eddies. 
We expect that when 
\begin{equation}
   D_{\|} \sim l_{\|}v_A
\end{equation} 
on a (parallel) scale 
\begin{equation}
  l_{c,\|} \sim D_{\|}/v_A \sim l_{mfp}v_{ptl}/v_A,
   \label{eq_lcpar}
\end{equation}
random walk behavior is fully recovered.
Here $l_{mfp}$ is the mean free path by pitch-angle scattering.
Therefore, the momentum diffusion coefficient is largest for eddies
with the parallel size $\sim l_{c,\|}$.
Note that  $t_{diff} \sim t_{eddy}$ on the scale $l_{c,\|}$,
where $t_{diff}$ is the diffusion time and $t_{eddy}$ is the eddy
turnover time.
On this scale, 
the change in momentum during one eddy turnover time is 
\begin{equation}
  \Delta p \sim \frac{ dp }{ dt } \Delta t 
           \sim p \left( \frac{ v_{l,slow} }{ l_{\|} } \right)
                  \left( \frac{ l_{\|} }{ v_A } \right)
           \sim p \frac{ v_{l,slow} }{ v_A },
\end{equation}
where we used 
$dp/dt \sim p\nabla \cdot {\bf v}_{l,slow} \sim pv_{l,slow}/l_{\|}$
and $\Delta t \sim l_{\|}/v_A$ (see Appendix A).
The net momentum diffusion coefficient is 
similar to 
eq. (\ref{eq_main_l}):
\begin{eqnarray}
   D_p^{slow}
   \sim (p^2 v_{l,slow}^2/ v_Al_{\|}) \label{eq_111} \\
    \sim \frac{ p^2 v_A }{ L } \left( \frac{ v_{L,slow} }{ v_A }
                                  \right)^2,
  \label{eq_d_p_lc}
\end{eqnarray}
where we used $v_{l, slow} \sim v_{L,slow} (l/L)^{1/3}$,
and $l_{\|} \sim L^{1/3}l^{2/3}$ (GS95).
Note that this result becomes identical to that in equation 
(\ref{eq_ptuskin_re2})
when we use the fact $D_{\|}\sim l_{\|}v_A$ 
(or, $t_{l,eddy}\sim t_{l,diff}$). 
One can still characterize the particle acceleration arising from slow modes
 at large scales using eq. (\ref{eq_ptuskin_re2}), but this contribution
will be sub-dominant to that by smaller scales given by eq. (\ref{eq_d_p_lc}).

To summarize, the momentum diffusion coefficient in slow diffusion limit
depends on turbulent diffusion.
We can write
\begin{equation}
  D_p^{slow}
    \sim \frac{ p^2 v_A }{ L } \left( \frac{ v_{L,slow} }{ v_A }
                                  \right)^2 Q_{TD},
  \label{eq_qtd}
\end{equation}
where $Q_{TD} \sim 1$ when turbulent diffusion is slow 
(see eq. (\ref{eq_d_p_lc})) and
$Q_{TD}\sim \ln(Lv_A/D_{\|})$ when it is fast 
(see eq. (\ref{eq_main_tot})).

\subsection{Slow and fast diffusion in high-$\beta$ limit}
So far in this section, we considered acceleration by 
slow modes in low-$\beta$ plasmas:
$\beta \leq O(1)$. 
In this case, 
$\nabla \cdot v_{l,slow} \sim v_{l,slow}/l_{\|}$.
However, when $\beta \rightarrow \infty$, we have
\begin{equation}
   \nabla \cdot v_{l,slow} \sim v_{l,slow}/(\beta l_{\|})
\end{equation}
(see Appendix A).
We can use the same $\Delta t$: $\Delta t \sim l_{\|}/v_A$.
Therefore, in high-$\beta$ limit, eq. (\ref{eq_sm_fd1})
becomes
\begin{equation}
  D_p^{slow} \sim \frac{ p^2 v_A }{ L } 
            \frac{ Lv_A }{ D }
            \left( \frac{ v_L }{ v_A } \right)^2   \beta^{-2}
   \label{eq_ptuskin_fast2_hb}
\end{equation}
for {\it fast diffusion limit}
and
eq. (\ref{eq_d_p_lc}) becomes
\begin{equation}
   D_p^{slow}
    \sim \frac{ p^2 v_A }{ L } \left( \frac{ v_{L,slow} }{ v_A }
                                  \right)^2 \beta^{-2} Q_{TD}
  \label{eq_d_p_lc_hb}
\label{41}
\end{equation}
for {\it slow diffusion limit}.
See eq. (\ref{eq_qtd}) for definition of $Q_{TD}$.
Evidently due to the $\beta^{-2}$ factor the acceleration by 
slow modes in high beta plasma is suppressed compared to the
case of low beta plasma.

\section{Acceleration by compressible motions: fast modes}
It is easy to see that arguments in the previous section 
(\S\ref{sect_slow})  are
applicable to  the acceleration of cosmic rays by fast modes,
which are shown to be isotropic (Cho \& Lazarian 2002).
As we will see, fast modes, which are essentially compressions
of magnetic field, can be very efficient.
Especially in low-$\beta$ plasmas, compression occurs in the
directions perpendicular to the magnetic field and
there is no adiabatic loss/gain in parallel directions.
Therefore the acceleration mechanism in \S \ref{sect_21} will be
most relevant.
Here we consider acceleration by fast modes in moderately compressible 
turbulence, in which shock formation is marginal.

\subsection{Fast diffusion limit}
In fast diffusion limit the acceleration by fast and slow modes
is very similar. Indeed,
consider $D_{\|}\gg Lc_f$, where $c_f$ is the propagation speed
of the fast wave and
depends on the plasma $\beta$: $c_f \sim v_A$ for
low-$\beta$ and $c_f \sim a$ (=sound speed) for high-$\beta$.
In this limit, particle diffusion is so fast that particles diffuse out
of even largest fast mode wave packets, or eddies, before the eddies
complete one oscillation. When a particle diffuse through an isotropic 
fast-mode eddy of size $l$, the change in momentum is 
\begin{equation}
   \Delta p \sim \frac{ dp }{ dt }\Delta t 
            \sim \frac{ pv_{l,fast} }{ l }
                 \frac{ l^2 }{ D_{\|} }
   =p \frac{ v_{l,fast}l }{ D_{\|} }, 
\end{equation}
where we used $\nabla \cdot {\bf v}_{l,fast} \sim v_{l,fast}/l$ and 
$\Delta t \sim l^2/D_{\|}$.
Therefore,
\begin{equation}
   D_{p,l} \sim (\Delta p)^2 /\Delta t 
   \sim p^2 v_{l,fast}^2/D_{\|},    \label{eq_fm_fd_l}
\end{equation}
which is most efficient on the outer scale $L$.\footnote{
   This statement remains true even the case particles re-enter
   the same eddy multiple times. 
   As in Chandran \& Maron (2004b), the re-entry factor is given by 
   $N\sim \min\{\sqrt{ D_{\|}\tau_{l,rand}}/l, M_{l,fast}\}$, where
   $\tau_{l,rand}$ is the eddy randomization 
   time of the fast modes on scale $l$. 
   Although $\tau_{l,rand}$ is uncertain,
   it is certain that $\tau_{l,rand} \geq t_{l,eddy}$ because
   fast modes cascade may be slower than Alfvenic cascade, where
   $t_{l,eddy}$ is the eddy turnover time of Alfvenic turbulence.
   In this case, the first term inside the parenthesis scales with
   $l^q$, where $q \leq l^{-2/3}$.
   The factor $M_{l,fast}$ is roughly $\sim z_s/l$,
   where $z_s$ is the distance along magnetic field lines over which two
   adjacent magnetic field lines get separated by the distance $l$.
   In the presence of strong Alfvenic turbulence, 
   $z_s \sim L^{1/3}l^{2/3}$. Therefore, the second term 
   inside the parenthesis scales with
   $l^{-1/3}$.
   Therefore, the second term is smaller and $N\propto l^{-1/3}$.
   The product of $D_p$ in eq. (\ref{eq_fm_fd_l}) and 
   $N$  scales with $v_{l,fast}^2 l^{-1/3}$.
   If fast modes follow scaling of acoustic turbulence,
   $v_{l,fast}^2 \propto l^{1/2}$.
   If fast modes follow Kolmogorov scaling,
   $v_{l,fast}^2 \propto l^{2/3}$.
   For either case, $v_{l,fast}^2 l^{-1/3}$ decreases as $l$ decreases.
   Therefore, the outer scale renders largest contribution
   to $D_p$.
}

 Setting $l=L$, we obtain
\begin{equation}
   D_p^{fast} \sim  
    \frac{p^2 v_A}{L}
    \frac{ Lc_f }{ D_{\|} }   
    \frac{ v_A }{ c_f } 
   \left( \frac{ v_{L,fast} }{ v_A } \right)^2.
   \label{eq_fastm_fastd}
\end{equation}
If $v_{L,fast} \sim v_A$ on the outer scale of turbulence
and when $D_{\|} > Lc_f$,
we have
\begin{equation}
   D_p^{fast} \sim \frac{ p^2 v_A }{ L }
                   \left( \frac{ Lc_f }{ D_{\|} }    \right)
     < \frac{ p^2 v_A }{ L }
\label{45}
\end{equation}
for low-$\beta$ plasmas and
\begin{equation}
   D_p^{fast} \sim \frac{ p^2 v_A }{ L } 
                   \left( \frac{ Lc_f }{ D_{\|} } \right)  \beta^{-1/2}
       < \frac{ p^2 v_A }{ L } \beta^{-1/2}
\label{46}
\end{equation}
for high-$\beta$ plasmas.
Note that the result in eq. (\ref{eq_fastm_fastd}) is 
very similar to that
for Ptuskin's acoustic turbulence (eq. (\ref{eq_ptuskin_1app})).
This is not so surprising because 
fast modes in MHD are similar to 
weakly interacting isotropic acoustic waves (see Cho \& Lazarian 2002). 

Similarly to the case of slow modes, factor $N$ that characterizes the
probability of the return of the particles back to the accelerating eddy
is applicable. Therefore eqs. (\ref{eq_fastm_fastd}), (\ref{45}) and (\ref{46})
should be multiplied by this factor. With this addition 
eq. (\ref{eq_fastm_fastd}) gets analogous to the eq. (\ref{27}). Although
$\beta$ factors appear both for slow and fast modes in eqs. ( \ref{eq_ptuskin_fast2_hb}) and (\ref{46}), respectively, they have different origins. 
For slow modes this corresponds to the suppression of efficiency, while
for fast modes the $\beta$-factor reflects the fact that the injection 
velocity is $v_A$. 

\subsection{Slow diffusion limit}
In this subsection, we consider $d_{min}^{}c_f \ll D_{\|}\ll L c_f$, 
where
$d_{min}$ is either the size of eddies at the dissipation scale
or the mean free path of the cosmic rays, 
whichever is larger.
Particle diffusion time is smaller than the {\it wave period} at the 
outer scale $L$ and larger than that on the scale $\sim d_{min}$.

In principle, the diffusion coefficient for motions
at a scale $l >l_c$, where 
\begin{equation}
     l_c \sim l_{mfp}v_{ptl}/c_f 
               \sim \left\{ \begin{array}{ll}
                           l_{c,\|} &    \mbox{low-$\beta$} \\
                           l_{c,\|} \beta^{-1/2} &   \mbox{high-$\beta$}
                           \end{array}
               \right.
\end{equation} 
(see eq. (\ref{eq_lcpar}) for $l_{c,\|}$),
can be obtained
using the approach in \S\ref{sect_slow} assuming that the coherence
time for a diffusing particle is equal to the period of
the fast wave, rather than the cascading time. 

When $D_{\|} \ll Lc_f$, we have $l_c \sim D_{\|}/c_f \ll L$.
However, we do not need to calculate $D_p$ for all scales between
$l_c$ and $L$.
As we will see later in eq. (\ref{eq_fast_vf}), the value of
$D_p$ is largest on the smallest scale $l_c$.
Therefore it suffices to calculate $D_p$ on the scale of $\sim l_c$.

The momentum change $\Delta p$
during the diffusion time, which is the same as the wave period 
on the scale of $l_c$,  is
\begin{equation}
 \Delta p \sim p (v_{l,fast}/l_c)(l_c/c_f).
\end{equation}
The momentum diffusion coefficient, $D_p$, is
\begin{eqnarray}
   D_{p,l}^{fast} \sim (\Delta p)^2/\Delta t  \nonumber \\
   \sim (p \frac{ v_{l,fast} }{ l_c } \frac{ l_c }{ c_f } )^2 
     / (l_c/c_f)
   \sim (p^2 v_{l,fast}^2/c_f l_c) \label{eq_fast_vf} \\
   \sim (p^2 v_{L,fast}^2/c_f l_c) (l_c/L)^{2m}  \\
   \sim (p^2 v_{L,fast}^2/c_f L) (l_c/L)^{2m-1}  \\
   \sim \frac{p^2 v_A}{L} \frac{v_{L,fast}^2}{v_A^2}\frac{v_A}{c_f}
        (l_c/L)^{2m-1},
\label{52}
\end{eqnarray}
where we %
assumed $v_{l,fast}\sim v_{L,fast} (l/L)^m$.
Note that $m=1/3$ for Kolmogorov turbulence and $m=1/4$
for acoustic turbulence (see Cho \& Lazarian 2002).

The difference between the slow diffusion limit for fast and slow modes
is evident from the comparison of eq. (\ref{52}) 
and eq. (\ref{eq_111}).   
The combination that enters eq. (\ref{52})  $v_{l,fast}^2/l$ 
increases as $l$ decreases, while the combination that enters  eq. (\ref{eq_111})
$v_{l,slow}^2/l_{\|}$ is independent of $l$.
Consequently, the rate of turbulent diffusion does not change
 much for fast modes.
Indeed, 
when turbulent diffusion is fast, we do not have suppression of random
walk for large scale eddies. Therefore, we may need to consider
all scale eddies between $L$ and $l_c$.
However, this does not really 
matter, if most contribution comes from the scales near the
scale $l_c$ for which the suppression is marginal even for
slow diffusion limit.

For low-$\beta$ plasmas, $c_f \sim v_A$ and fast modes give 
\begin{eqnarray}
   D_{p}^{fast} 
  \sim \frac{ p^2 v_A }{ L } \left( \frac{ v_{L,fast} }{ v_A } 
                                  \right)^2
                                  \left( \frac{ L }{ l_c }
                                  \right)^{1-2m}  \nonumber \\
 \sim \frac{ p^2 v_A }{ L } \left( \frac{ v_{L,fast} }{ v_A } 
                                  \right)^2
                                  \left( \frac{ Lv_A }{ l_{mfp}v_{ptl} }
                                  \right)^{1-2m},
   \label{eq_fastm_slowd_lb}
\end{eqnarray}
which is larger than that of slow modes if $l_c \ll L$.
Note that $1-2m > 0$ for $m< 1/2$.
For high-$\beta$ plasmas, $c_f \sim a$ (=sound speed) and fast modes give 
\begin{eqnarray}
   D_p^{fast} 
  \sim \frac{ p^2 v_A }{ L } \left( \frac{ v_{L,fast} }{ v_A } 
                                  \right)^2
                                  \beta^{-1/2}
                                  \left( \frac{ L }{ l_c }
                                  \right)^{1-2m} \nonumber \\
  \sim \frac{ p^2 v_A }{ L } \left( \frac{ v_{L,fast} }{ v_A } 
                                  \right)^2
                                  \beta^{-m}
                                  \left( \frac{ Lv_A }{ l_{mfp}v_{ptl} }
                                  \right)^{1-2m},
     \label{eq_fastm_slowd_hb}
\end{eqnarray}
which is larger than that of slow modes if $l_c \ll L$.

\section{Acceleration in Compressible highly supersonic turbulence}

When turbulence is highly supersonic, the treatment of
$\nabla \cdot {\bf v}$ should be different.
In subsonic case, when we consider motion on scale $l$,  the
general treatment is that
$\Delta p \sim (dp/dt)\Delta t \sim (\nabla \cdot {\bf v}) \Delta t
    \sim \Delta t v_l/l \sim$ constant, where we used 
$\Delta t \sim l/v_l$.
However, in the supersonic case, shocks compresses gas and 
form high density
regions (see, for example, Padoan, Nordlund, \& Jones 1997;
Padoan \& Nordlund 1999; Beresnyak, Lazarian \& Cho 2005; Kim \& Ryu 2005). 

The maximum compression for isothermal hydrodynamic turbulence scales as
$\sim M_s^2$, where $M_s$ is the 
sonic Mach number (Padoan et al. 1997).    
   We can derive this by equating turbulence pressure, 
    $\sim \bar{\rho} v_L^2$, and gas pressure in the high density
   regions, $\rho_{max} c_s^2$, where $\bar{\rho}$ is
   the average density and $\rho_{max}$ is the density of the 
   compressed regions. Here we assume isothermal gas.
   The result is 
\begin{equation}
   \rho_{max}/\bar{\rho} \sim v_L^2/c_s^2 = M_s^2.
\end{equation}
   This scaling is valid for hydrodynamic turbulence.

\begin{figure*}[h!t]
\includegraphics[width=.49\textwidth]{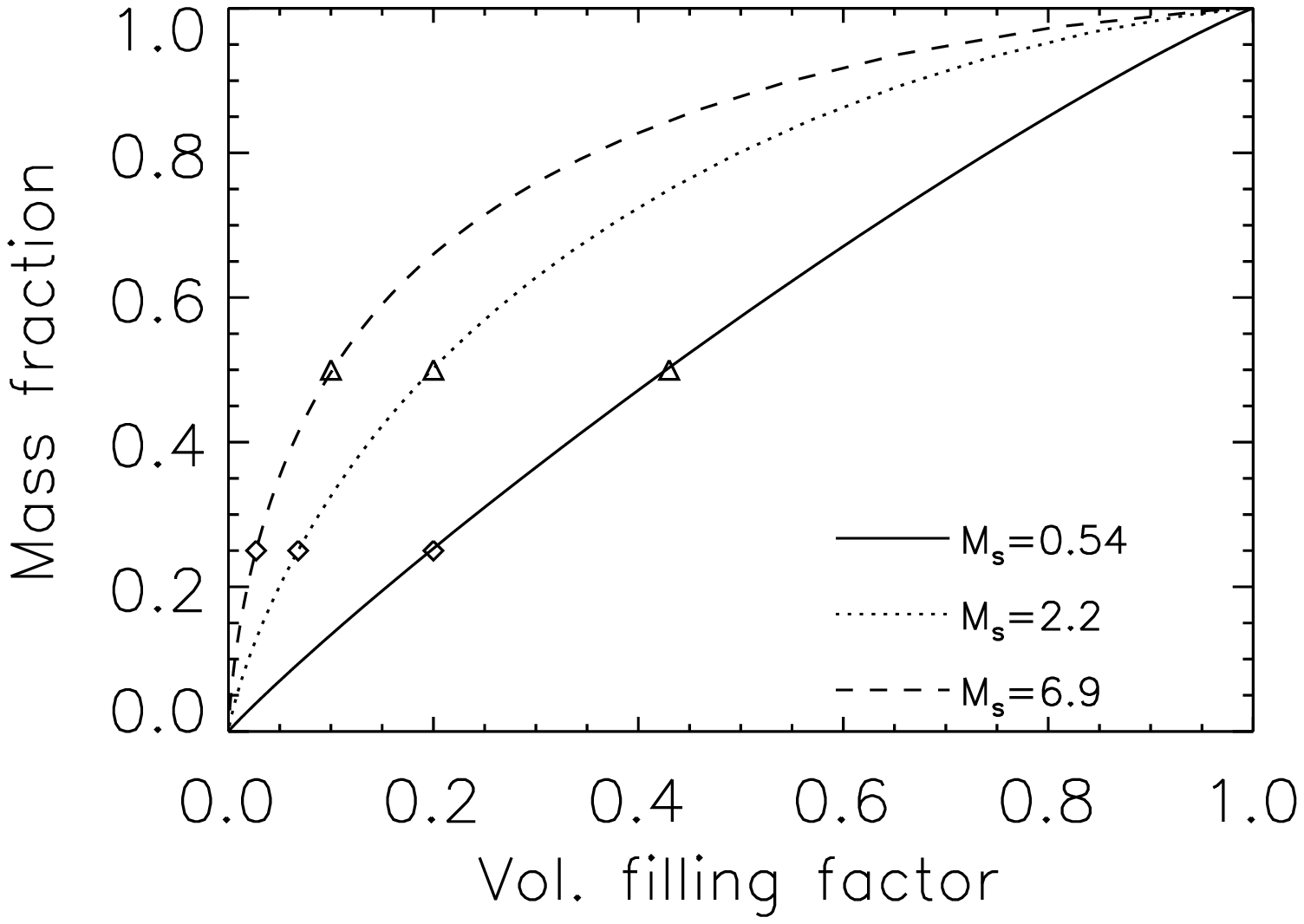}
\includegraphics[width=.49\textwidth]{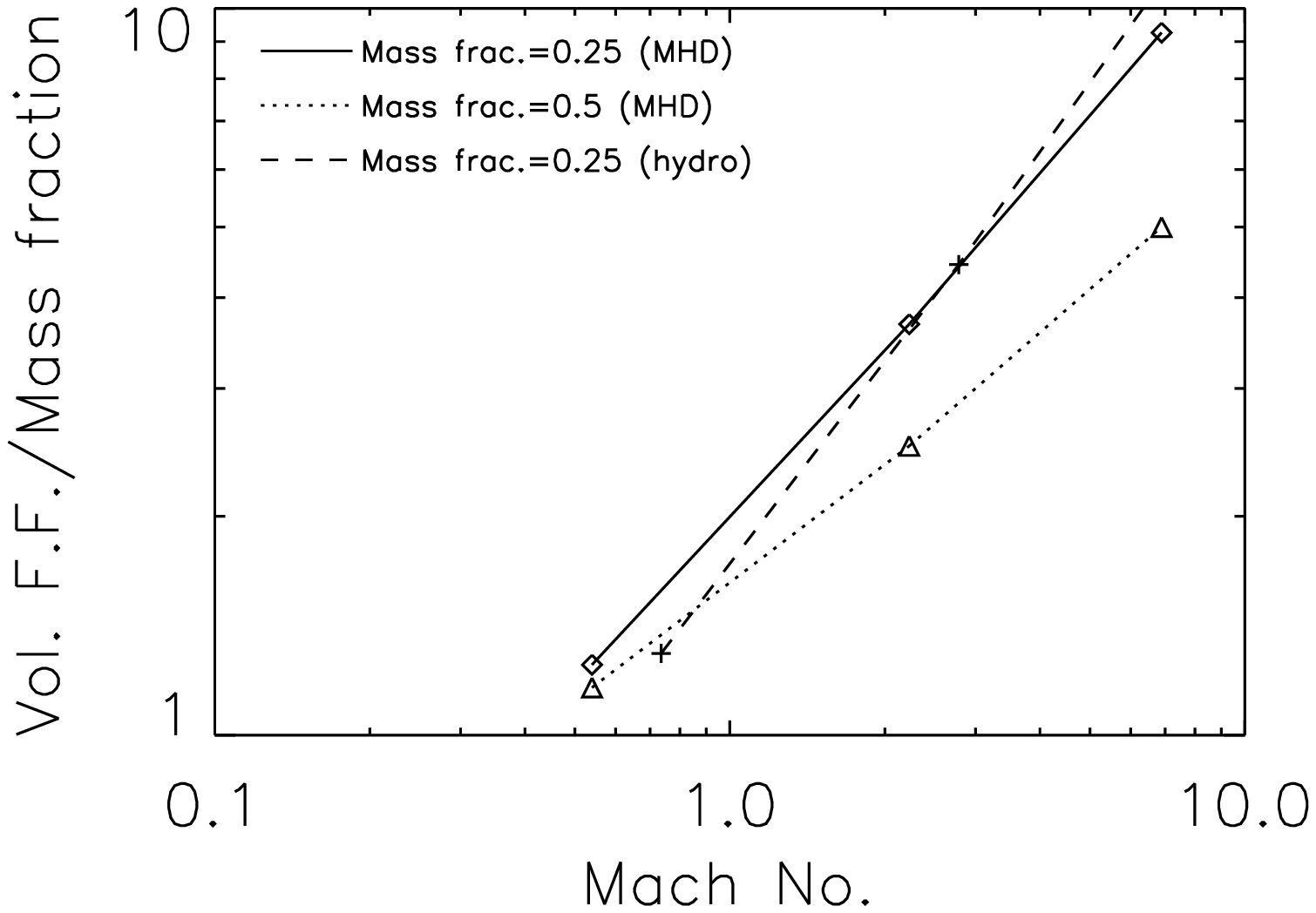}
\caption{ Compression rates in strongly magnetized 
   compressible MHD turbulence.
   ({\it Left}) When the sonic Mach number is higher, mass concentration
   is higher. For example, the triangle on the dotted curve,
   which is for $M_s \sim 2.2$, tells us that about half of the
   mass occupies 20\% of the volume.
   The diamond on the dashed curve says that
   25\% of mass fills 2.7\% of the volume.
  ({\it Right}) The compression factor may be defined by
  volume filling factor divided by the mass fraction for
  triangles and diamonds in the left panel.
  We plot these values as function of the sonic Mach number.
  Note that the solid line (MHD) and the dashed line (hydro)
  have similar slope. Although the solid line has a shallower slope,
  this figure roughly suggest that compression rate in 
  isothermal MHD medium shows a dependence on Mach number 
  similar to isothermal hydrodynamic one.
}
\label{fig_4}
\end{figure*}

It is not clear whether or not
such a simple scaling exists for MHD turbulence
(see discussions in Ostriker, Stone, \& Gammie 2001; 
Padoan \& Nordlund 2002).
   In MHD turbulence, magnetic pressure provides
   resistance for compression, which results in less compression.
   Since slow modes are almost 
   parallel to the local mean field directions, magnetic pressure
   may not provide a significant pressure when slow modes form shocks.
   Therefore, we expect that the maximum compression
   in MHD turbulence depends on the sonic Mach number
   even for strongly magnetized case. 

Our numerical simulations support this idea.
We use data cubes obtained from direct compressible MHD turbulence
simulations with $216^3$ cells. These simulation are described in CL03.
We used 3 data cubes with different sonic Mach numbers: 
$M_s \sim$ 0.5, 2.2, and 7.
In all simulations, we fix the r.m.s. velocity and the strength of the 
mean magnetic field. The Alfven speed of the mean magnetic field is 
very close to the r.m.s. velocity of turbulence.
To see how much compression is achieved, we plot
the relation between volume filling factor and fraction of mass contained inside the volume for high density regions 
(see left panel of Fig. \ref{fig_4}).
The figure shows that higher sonic Mach number fluids are compressed more,
which is not so trivial because the fluid is strongly magnetized.
Let use focus on diamond symbols in the figure.
For example, the diamond on the dashed curve in the left panel says that
   25\% of mass occupies 2.7\% of the total volume.

It is not clear how to derive exact scaling relations from  
Figure \ref{fig_4}.
However, we can derive at least two facts from the figure.
First, as we mentioned earlier, the compression rate
scales with  $M_s^{\gamma}$ even for strongly magnetized medium.
Second, 
the right panel of Fig. \ref{fig_4} show that 
the compression rates in MHD and hydrodynamic turbulence  scale very similarly. 
The dashed line in the right panel is for hydrodynamic turbulence.
Comparing the solid line and the dashed line on the right panel,
we note that the slopes are very similar.
It is natural to assume that the compression rate in hydrodynamic
turbulence scales with $M_s^{2}$.
Therefore, we can say that the compression rate, defined as the volume filling after compression 
divided by that before compression,in MHD
scales with  $M_s^{\gamma}$, where $\gamma$ is
slightly smaller than 2.

Here we assume that the maximum compression in a strongly magnetized
MHD turbulence
scales with $M_s^{\gamma}$. 
We also assume that spatial diffusion of particle is slow.
Then we have
$\Delta p /p \sim \rho_{max}/\bar{\rho} \sim M_s^{\gamma}$. 
The momentum diffusion coefficient is
\begin{eqnarray}
   D_p^{supersonic} 
   \sim (\Delta p)^2/\Delta t \sim p^2 M_s^{2\gamma} (v_L/L),
                    \nonumber \\
       \sim \frac{ p^2 v_A }{ L }
            \frac{ v_L }{ v_A } M_s^{2\gamma},
      \label{eq_supersonic}
\end{eqnarray}
where we consider only the largest scales because smaller
scales are less efficient.
If $v_L \sim v_A$, this diffusion coefficient is $\sim M_s^{2\gamma}$ times
larger than the one in C03. 
Note that this is a second order Fermi process associated with shocks. Although
compression of a single eddy leads to the first order Fermi acceleration,
the first order effect cancels out after the particle encounters
many compression and expansions and only the second order effect
leads to increase of momentum through diffusion process 
(see Bykov \& Toptygin 1982).

In this section we have implicitly assumed that particle diffusion
time is similar to or larger than the eddy turnover time
on the outer scale.
In the fast diffusion  limit the our earlier estimates of the
turbulent acceleration are valid.  Therefore
the the momentum diffusion coefficient is
of the order $\sim p^2 v_L^2/D_{\|}$, if the particle does not
return back to the high density compression. If it returns before
the compression disappears this estimated should be multiplied  by
$N>1$.

\section{Acceleration by electric field arising from slow modes in
   incompressible limit}   \label{sect_2}

Let us consider the extreme case of high $\beta$ plasma, namely,
incompressible plasma. In the incompressible limit only two modes,
Alfven and slow (pseudo-Alfven) exist. Slow modes compress magnetic
field.  

\subsection{Betatron Acceleration Process}   \label{sect_21}

\begin{figure*}[h!t]
\includegraphics[width=.95\textwidth]{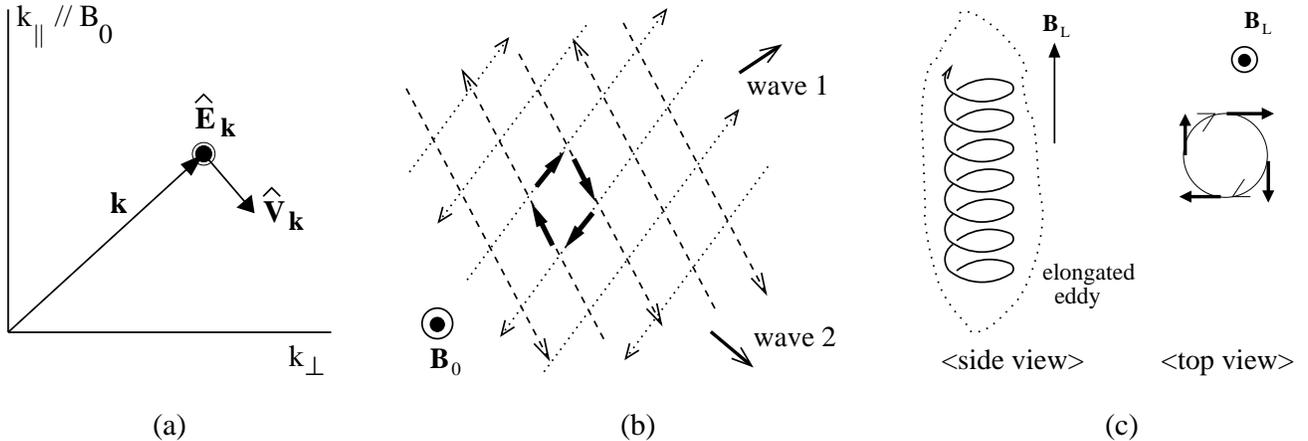}
\caption{Slow modes and non-zero $\nabla \times {\bf E}$.
(a) Direction of ${\bf E}$. Electric vector is perpendicular to both
    ${\bf B}_0$ and ${\bf k}$, which means that,
    when viewed from the top, electric field is parallel
    to the wave front. 
(b) Nonzero $\nabla \times {\bf E}$ by slow modes.
    The dashed and the dotted lines denote wave fronts of wave 1 and 2, 
    respectively. Arrows are the directions of electric field.
    Two slow waves can create non-zero  $\nabla \times {\bf E}$.
(c) Acceleration by an elongated slow mode eddy.
    The particle experiences constant acceleration in the eddy.
The integral $\int (\nabla \times {\bf E})\cdot d{\bf s}$ should not be
zero in the eddy.
}
  \label{fig:1}
\end{figure*}

In incompressible conducting fluid a slow wave
causes an oscillation occurring in the plane spanned by
${\bf B}_0$ and ${\bf k}$ vectors, where ${\bf B}_0$ is the mean magnetic
field, or more precisely \textit{local} mean magnetic field, 
and ${\bf k}$ is the wave vector.
Therefore, the electric vector 
${\bf E} \propto -{\bf v}_{\bf k} \times {\bf B}_0$ is perpendicular to
both ${\bf B}_0$ and ${\bf k}$. Here, ${\bf v}_{\bf k}$ is the
amplitude of the oscillation velocity of the pseudo-Alfven wave.
The electric field is parallel to the wave front (see Figure \ref{fig:1}).
Thus, slow modes can create non-zero $\nabla \times {\bf E}$
(see Figure \ref{fig:1}(b)).
When a particle gyrates around ${\bf B}_0$, the particle feels
electric field caused by turbulence.
When the electric field has non-zero curl,
the particle experiences either acceleration or deceleration.
An accelerating eddy\footnote{An eddy-wave duality in the description of
MHD turbulence is discussed in CL03.} is shown in Figure \ref{fig:1}(c).

Consider a particle with gyro radius $r_g$ moving
 through turbulent eddies
of (perpendicular) scale $l>r_g$.
Then the average electric field the particle feels during one
gyro rotation is
\begin{eqnarray}
  \bar{E}= \frac{1}{2\pi r_g} \oint {\bf E} \cdot d{\bf l} 
     = \frac{1}{2\pi r_g} \int (\nabla \times {\bf E})\cdot d{\bf s} 
     \nonumber \\
  = \frac{1}{2\pi r_gc} 
     \int ( {\bf B}_0 \nabla \cdot {\bf v}_l
           -{\bf B}_0 \cdot \nabla {\bf v}_l ) \cdot d{\bf s}, 
\label{eq_1} \\
  = -\frac{1}{2\pi r_gc} 
     \int ( {\bf B}_0 \cdot \nabla {\bf v}_l ) \cdot d{\bf s}, 
 \label{eq_2}
\end{eqnarray}
where $d{\bf l}$ is the infinitesimal line element vector,
      $d{\bf s}$ the infinitesimal area element vector, 
      ${\bf E}=-{\bf v}\times {\bf B}_0/c$, and
      $c$ the speed of light.
The area element vector $d{\bf s}$ is taken to be
parallel to ${\bf B}_0$.
Eq. (\ref{eq_1}) is valid for both incompressible and compressible
fluids. However, eq. (\ref{eq_2}) is valid only for incompressible
fluids because we used $\nabla \cdot {\bf v}=0$ when we derived it.
The integral is zero for Alfven modes because
${\bf v}_l \cdot d{\bf s}=0$.
For slow modes, the integral is approximately
\begin{equation}
 \int ({\bf B}_0 \cdot \nabla {\bf v}_{l}) \cdot d{\bf s} \approx
  B_0 v_{l\|} (\pi r_g^2)/l_{\|},
  \label{eq_3int}
\end{equation}
where $v_{l\|}$ is the velocity component of pseudo-Alfv\'en modes
parallel to ${\bf B}_0$.
Thus, we have
\begin{equation}
  \bar{E} \approx -B_0 v_{l\|} r_g/2l_{\|}c.
\label{electric}
\end{equation}
The amplitude of $\bar{E}$ is the largest\footnote{Note, that 
for particle acceleration not only the amplitude of $\bar{E}$ but also
its correlation time matters. The latter is larger for larger eddies.}
 at $l\sim r_g$. Indeed,
when $l>r_g$, since 
$v_{l\|} \sim v_l \propto l^{1/3}$ and 
$l_{\|} \propto l^{2/3}$ (GS95), 
$\bar{E} \propto l^{-1/3}$ and 
$\bar{E}$ increases with the decrease of $l$.
On the contrary,
when  $l<r_g$, the particle traverses many uncorrelated eddies 
during one gyro orbit. 
The value of $(\nabla \times {\bf E})_l$ in an 
eddy of size $l$ is
of order 
$(\nabla \times {\bf E})_l \sim {\bf B}_0 \cdot \nabla {\bf v}_l/c
\sim B_0 v_l/l_{\|}c \propto l^{-1/3}$. This value times
$\Delta s \sim \pi l^2$ is the total contribution to $\bar{E}$ by the eddy
(c.f.~eq. (\ref{eq_3int})). 
There are $\sim (r_g^2/l^2)^{1/2}$ uncorrelated eddies 
inside the particle's gyro orbit.
Therefore the area integration (cf.~eq. (\ref{eq_2})) 
of these small scale fluctuations
over a circle of radius $r_g>l$ yields
$
  \bar{E} \propto ( r_g^2/l^2 )^{1/2} (\pi l^2)l^{-1/3} \sim r_g l^{2/3}.
$
As the result, the $\bar{E}$ increases with $l$.

\subsection{Structure of electric field of slow modes}

In this section we use a data cube obtained from a direct 
incompressible MHD turbulence simulation with $256^3$ grid points 
(Cho \& Vishniac 2000)
to study the structure of ${\bf E}$.
The data cube clearly shows that MHD fluctuations do produce
non-zero $\nabla \times {\bf E}$.

\begin{figure*}[h!t]
\includegraphics[width=.47\textwidth]{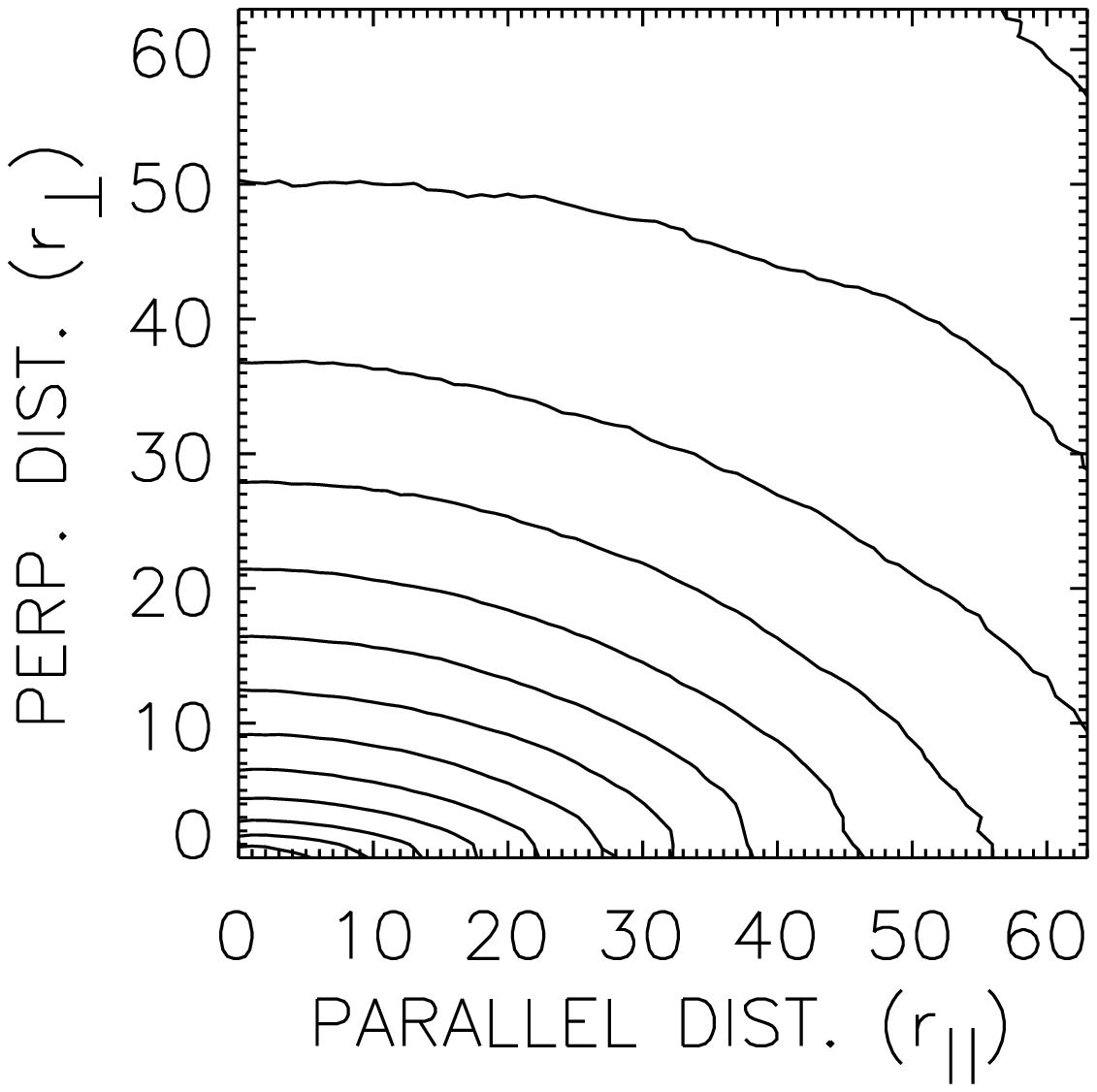} \hspace{5mm}
\includegraphics[width=.49\textwidth]{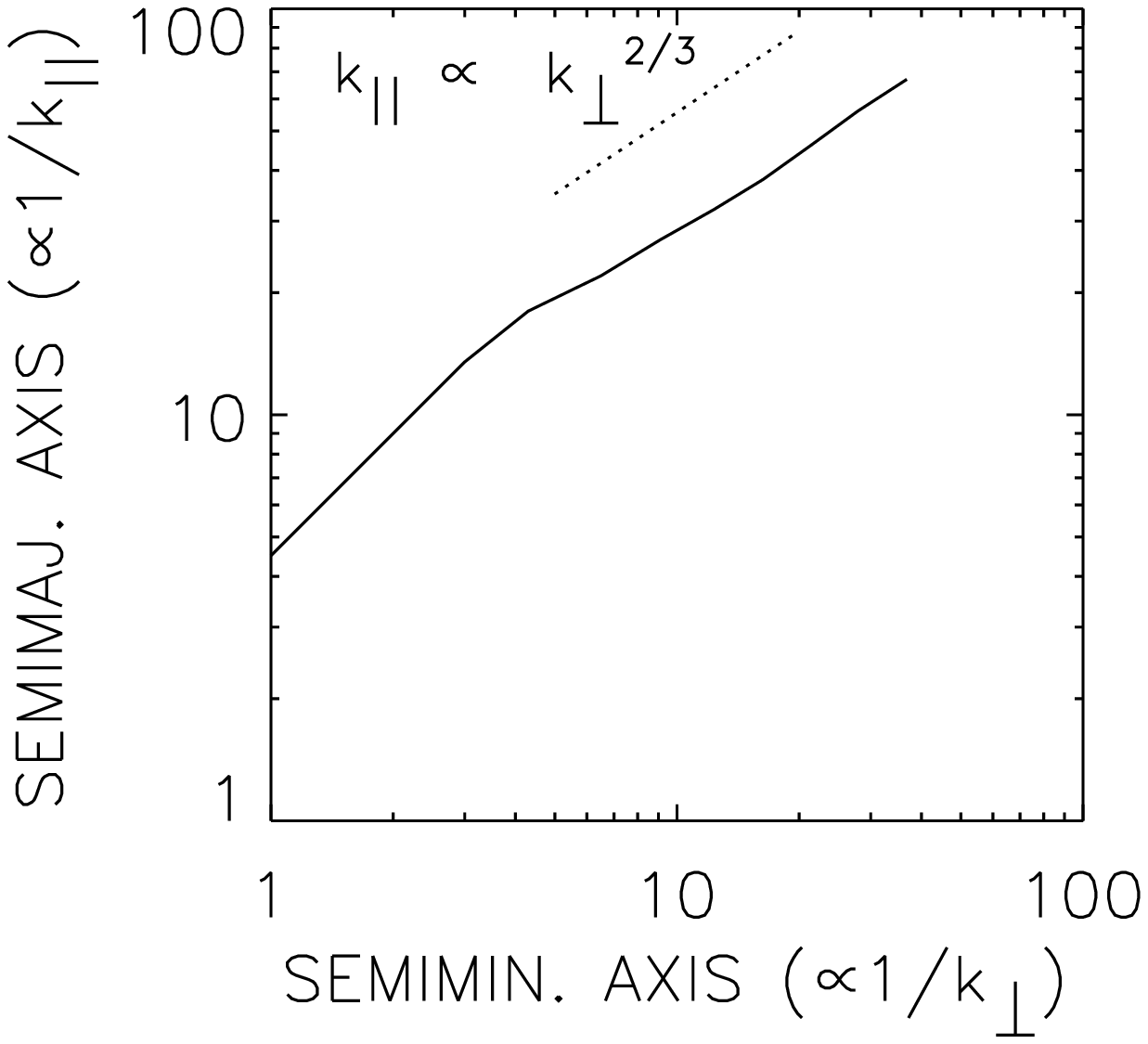}
\\
\begin{center}
(a)  \hspace{60mm} (b)
\end{center}
\caption{
  Anisotropy of ${\bf E}$ field.
  \textit{(a)} The second order structure function of ${\bf E}$.
  \textit{(b)} Scale-dependent anisotropy. Results are consistent with
      Goldreich-Sridhar type anisotropy: $k_{\|} \propto k_{\perp}^{2/3}$.
}
 \label{fig:2}
\end{figure*}

\begin{figure*}
\includegraphics[width=.48\textwidth]{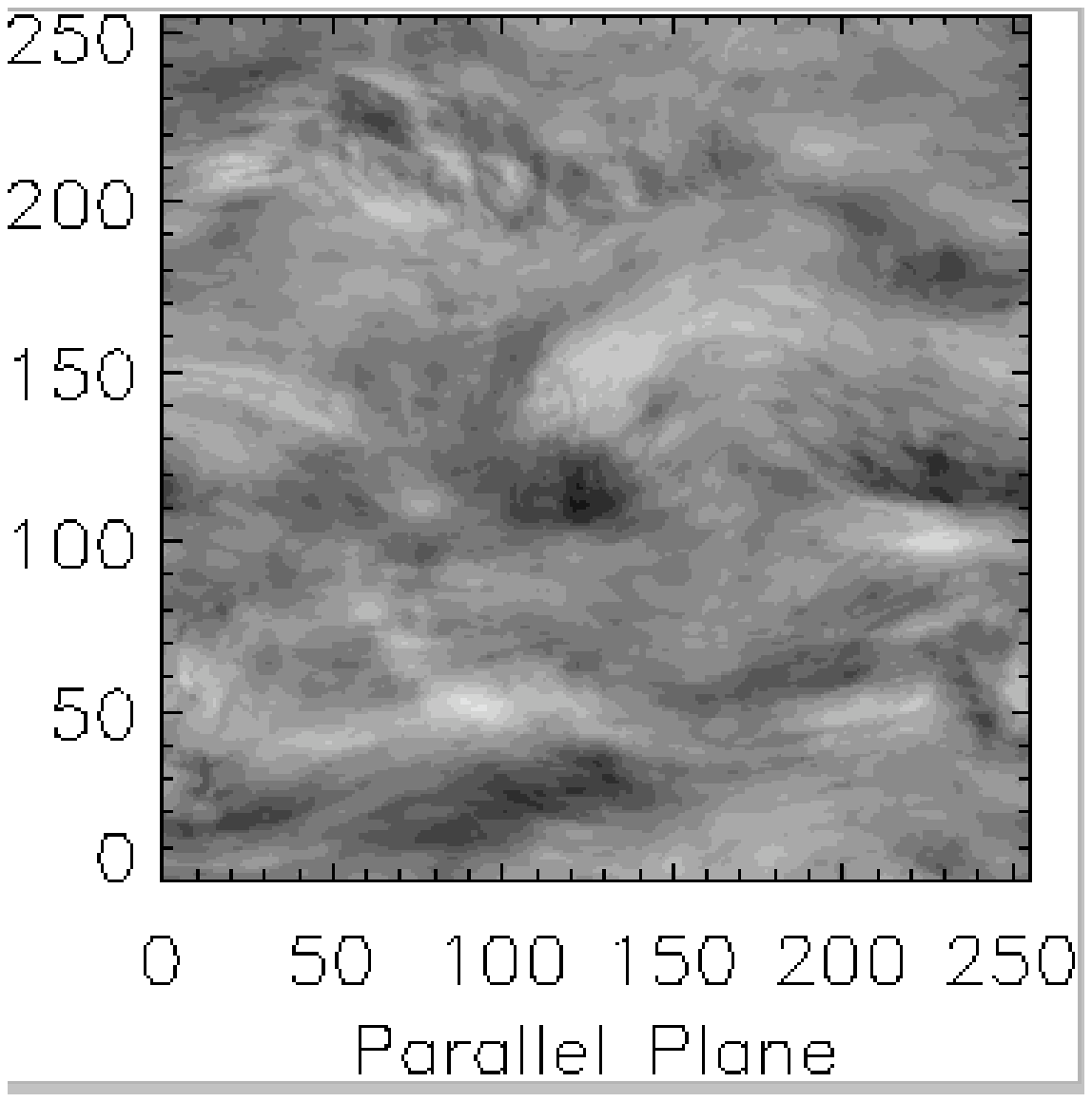}  
\includegraphics[width=.48\textwidth]{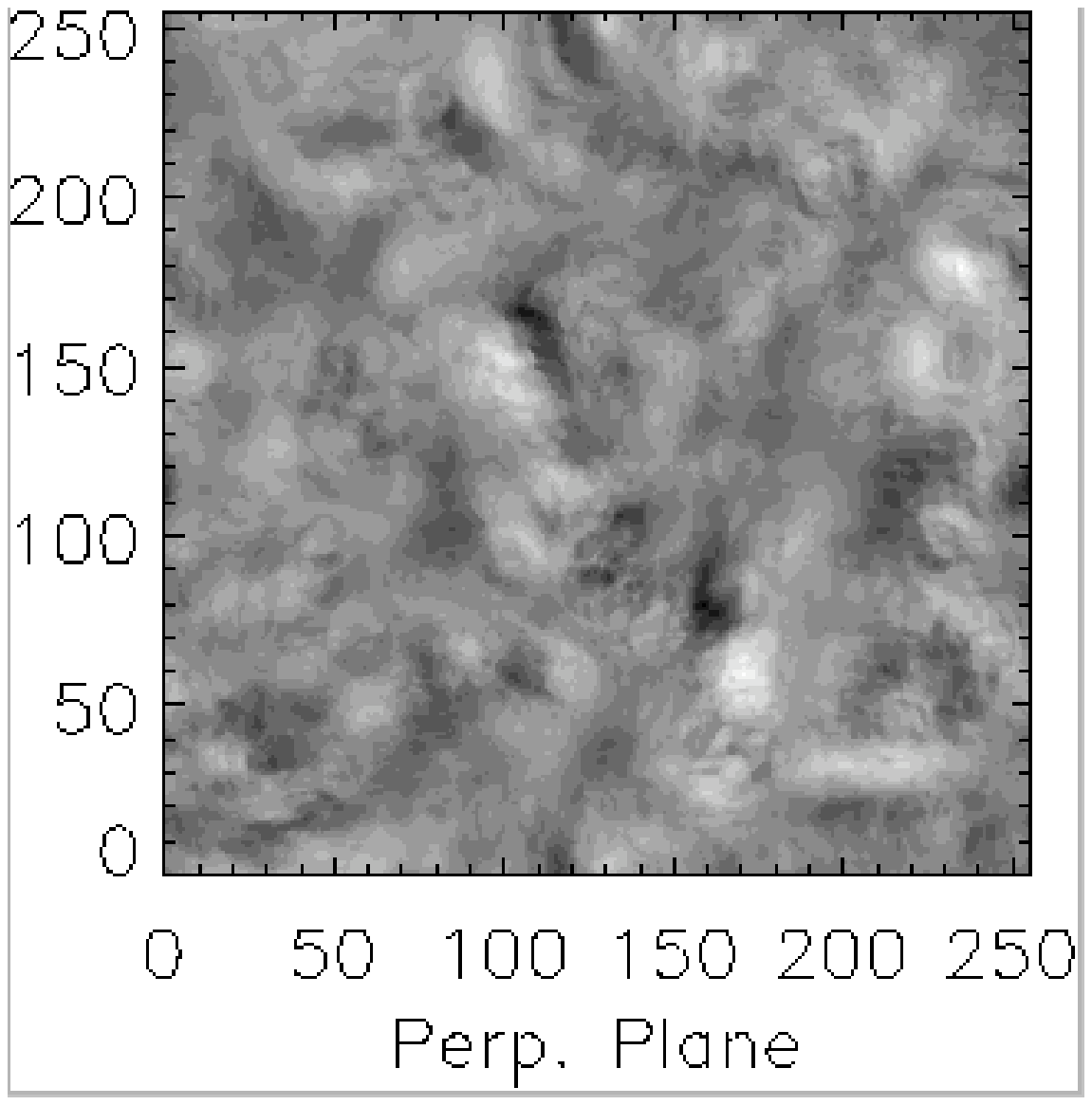} 
\\
\begin{center}
        (a)  \hspace{60mm} (b)  
\end{center}
\caption{Anisotropy of E field.
  Comparison of
    anisotropy of $\oint {\bf E}\cdot d{\bf l}
    =\int_{r_g} \nabla \times {\bf E} \cdot d{\bf s}$ parallel 
  (panel (a))
  and perpendicular (panel (b)) to magnetic field ($r_g$=8).
  The mean field ${\bf B}_0$ is along the horizontal axis in (a) and
  perpendicular to the plane in (b).
}
 \label{fig:3}
\end{figure*}

In Figure \ref{fig:2}(a), we plot contours of 
equal second order structure function
for electric field
in a local frame, which is aligned with the local mean magnetic field
${\bf B_L}$:
$
    \mbox{SF}_2(r_{\|},r_{\perp})=<|{\bf E}({\bf x}+{\bf r}) -
                 {\bf E}({\bf x})|^2>_{avg.~over~{\bf x}},
$
where ${\bf r}=r_{\|} {\hat {\bf r}}_{\|} +r_{\perp} {\hat {\bf r}}_{\perp}$
and ${\hat {\bf r}}_{\|}$ and ${\hat {\bf r}}_{\perp}$ are unit vectors
parallel and perpendicular to the local mean field ${\bf B_L}$, respectively
(see Cho \& Vishniac (2000) for a 
detailed discussion of the local frame).
          The contour plot clearly
          shows the existence of scale-dependent anisotropy:
          smaller eddies are more elongated.
By analyzing the relation between the semi-major axis 
($\sim l_{\|} \sim 1/k_{\|}$) and the semi-minor
axis ($\sim l_{\perp} \sim 1/k_{\perp}$) of the contours, 
we can obtain the 
relation between $k_{\|}$ and $k_{\perp}$.
Here $k_{\perp}$ and $k_{\|}$ are wave numbers perpendicular and
parallel to the mean magnetic field, respectively.
The result in Figure \ref{fig:2}(b) is consistent with the 
GS95 type anisotropy:
$
   k_{\|} \propto k_{\perp}^{2/3}.
$
In Figure \ref{fig:3}, we plot 
$\oint_{r_g} {\bf E}\cdot d{\bf l}$ for $r_g$=8 in grid units.
The particle orbit lies within the inertial range of turbulence,
which spans from $\sim 100$ grid units to a few grid units
in the simulation data.
We can see that structures are elongated along
the mean field direction. 

\subsection{Adiabatic gain/loss in parallel directions}   \label{sect_23}

The rate of momentum change by the
electric field is
\begin{equation}
   \frac{ dp_{\perp} }{ dt }=q\bar{E} \sim \frac{ q B_0 v_{l\|} r_g}{2l_{\|}c}
\sim \frac{p_{\bot}c}{r_g} \frac{v_{l\|} r_g}{2l_{\|}c} 
\sim \frac{p_{\bot} v_{l\|}}{2l_{\|}},
                          \label{eq_dpdt}
\end{equation}
where we used $qB_0=p_{\bot}c/r_g$, where $p_{\bot}$ is the particle momentum
perpendicular to magnetic field.
Since the integrand in eq. (\ref{eq_1}) is
$ {\bf B}_0 \nabla_{\perp} \cdot {\bf v}_{l\perp}$,
where '$\perp$' denotes the component perpendicular to ${\bf B}_0$, 
we can rewrite eq. (\ref{eq_dpdt}) as follows:
\begin{equation}
    dp_{\perp}/dt 
    = -p_{\perp} \nabla_{\perp} \cdot {\bf v}_{\perp} /2.
\label{general}
\end{equation}
This means that the acceleration by electric field is achieved
when large scale turbulence compresses magnetic field
in perpendicular directions.
The change of perpendicular component of energy is
\begin{equation}
   \frac{ \dot{E}^K_{\perp} }{ E^K_{\perp} }
   = 2 \frac{ \dot{p}_{\perp} }{ p_{\perp} }
   = \frac{ v_{l\|} }{ l_{\|} },  \label{eq_eperp}
\end{equation}
where $E^K$ is the kinetic energy and 
we consider non-relativistic case for simplicity. 

When we consider incompressible fluids, compression in 
perpendicular directions causes expansion in parallel 
directions,  which in turn results in adiabatic energy loss.
The rate of momentum change
by the parallel motions alone is
\begin{equation}
   \frac{ \dot{p}_{\|} }{ p_{\|} }
  =-\frac{ v_{l\|} }{ l_{\|} }.
\end{equation}
Therefore, the change of parallel component of the energy is
\begin{equation}
    \frac{ \dot{E}^K_{\|} }{ E^K_{\|} }
   = 2 \frac{ \dot{p}_{\|} }{ p_{\|} }
   = -2 \frac{ v_{l\|} }{ l_{\|} }.  \label{eq_epar}
\end{equation}
{}From equations (\ref{eq_eperp}) and (\ref{eq_epar}), we get
\begin{equation}
   \frac{ d(E^K_{\perp} + E^K_{\|}) }{ dt } 
   = (E^K_{\perp} - 2E^K_{\|}) \frac{v_{l\|}}{l_{\|}} = 0 \label{eq_etot}
\label{10}
\end{equation}
when particle distribution is isotropic (i.e. $E^K_{\perp} = 2E^K_{\|}$).

Therefore, 
if the particle distribution is isotropic, 
the second term in eq. (\ref{eq_1}), 
or the term in eq. (\ref{eq_2}), is canceled by adiabatic gain/loss in 
parallel directions and only the $\nabla \cdot {\bf v}_l$ term 
in eq. (\ref{eq_1}) survives, the result of which is equivalent to
the hydrodynamic approach of P88 that we discuss in the
next section, i.e.
\begin{equation}
   dp/dt \approx -p \nabla \cdot {\bf v}.   \label{eq_hydro}
\end{equation}

However, it is clear from eq.~(\ref{10}) that an acceleration instability is
present when betatron acceleration acts. Indeed, as electric field increases 
$E^K_{\bot}$, 
the gain of the energy gets positive. As processes of scattering
tend to isotropize particle velocities, we may assume that this 
instability acts on the time scale of randomization of a particle. This time
can be of the order of mean free path divided by the particle velocity and multiplied by a factor larger than unity (see discussion in Jokipii 1968). 
For our simplified treatment we conservatively assume that this factor is
order of unity. 
In other words we claim that,
on the mean free path scale of the particles, non-resonant effects are
important for particle motions and, therefore, the hydrodynamic
approach of P88  fails. 

Consider an eddy  a few times smaller than the mean free path.
Let this scale be $l^{\prime}$. 
Suppose that the eddy is compressed in the perpendicular directions when
a particle is back-scattered in the eddy. 
During the back-scattering, the particle's pitch angle is obviously large: around $\sim 90^{\circ}$.
Suppose that the particle stays in this large pitch angle state
for $\Delta t$.
Later, the particle will have smaller pitch angles.
On average, the particle will be in the small pitch angle state 
for $\Delta t$.
While the particle is in the small pitch angle state, the particle
will travel more distance because the parallel velocity is higher.
Therefore, the particle will pass through several eddies of the
scale $l^{\prime}$, 
the acceleration/deceleration by such eddies will roughly
cancel out. As a result, the motions of
eddies that contain the turning points are more important. 
Since such motions
are mutually independent,
we can have diffusion of momentum.

Since large scale eddies do not provide net acceleration when
particle distribution is isotropic, 
we do not consider large scale eddies in the incompressible
limit.
In the next subsection, we consider acceleration by
eddies on the mean free path scale.

\subsection{Estimates of $D_p$}

When particle distribution is isotropic, we do not have net
acceleration (see eq. (\ref{eq_etot})) by large scale eddies.
However, as we discussed in the previous subsection, 
equation
(\ref{eq_etot}) is no longer valid on the mean free path scale.

Here, we use properties of strong MHD turbulence, 
the simplest case of which
is that turbulence is isotropic on the
energy injection scale $L$ and that $B_0 \sim b_L$, where $b_L$ is the rms fluctuating field.
For eddies whose parallel size is comparable to the mean free path
by pitch-angle scattering, 
$l_{mfp}$, 
particles are accelerated for $\Delta t \sim l_{mfp}/v_{ptl}$, where 
$v_{ptl}$ is particle velocity. 
Note that in the absence of fast modes
the mean free path may be very large as scattering by Alfven and slow modes is marginal (Chandran 2000; YL02). 
The momentum change over the particle's eddy crossing time, 
$\Delta t \sim l_{\|}/v_{ptl}$, is 
\begin{equation}
 \Delta p \sim (dp_{\perp}/dt)\Delta t 
   \sim p_{\bot} v_{l\|}/2v_{ptl},
   \label{eq_Dp_main}
\end{equation}
where we used 
the acceleration that arises from magnetic field compression (see eq. (\ref{general})):
$dp/dt 
  \sim pv_{l,slow}/l_{\|}$ (see Appendix).
Therefore, 
\begin{eqnarray}
  D_{p}^{incomp} \sim (\Delta p)^2/\Delta t 
   \sim (p^2 v_{L,slow}^2/6v_{ptl} L),  \nonumber \\
   \sim \frac{ p^2 v_A }{ L } \left (\frac{ v_{L,slow} }{ v_A }\right)^2
       \left( \frac{ v_{A} }{ 6v_{ptl} } \right),
\label{Dpl}
\end{eqnarray}
where we used $v_{l\|} \sim v_{L,slow} (l/L)^{1/3}$ and $l_{\|} \sim L^{1/3}l^{2/3}$.
For $v_{L,slow}\sim v_A$, this is $\sim v_A/v_{ptl}$ times
smaller than eq. (30) in 
C03 (see Appendix A).
The ratio  $v_A/v_{ptl}$ is small for the interstellar medium, since
$v_A$ is a few km/sec and $v_{ptl}$ can be as large as the speed of light.
However, for a typical solar coronal loop, this ratio is $\sim 1/30$
since $v_A \sim 10^4$ km/sec (see C03).
This mechanism is more efficient than particle acceleration
by large scale compressive motions by slow modes (\S\ref{sect_slow})
in the high-$\beta$ limit, where the efficiency of the latter goes to zero.

As the interaction happens over the time of ballistic passage of a particle through an
eddy it is fast for all the physically interesting cases. 
Random walk in terms
of the applied field may also be applicable, if turbulence scale is less than the 
mean free path of a particle. This would only decrease the efficiency of the mechanism,
however.

\section{Acceleration by Weak Turbulence}

Although we believe that for most of astrophysical, e.g. interstellar (see
Armstrong, Rickett \& Spangler 1995),
 conditions
the GS95 scaling of slow modes presents the best-known fit,
{\it weak} turbulence (Galtier et al. 2000) may also arise in astrophysical
situations (Saur et al. 2002).
In weak turbulence, $v_L \ll v_A$ and turbulent diffusion is very slow.
Particle acceleration by anisotropic weak turbulence
is useful, as
MHD perturbations at large scales may evolve initially along weak
turbulent cascade before turning into the strong cascade (see a
discussion in Cho, Lazarian \& Vishniac 2003). 
In this section, we consider particle acceleration in
weak turbulence. Since large-scale Alfvenic motions 
do not accelerate particles, we discuss slow and fast modes.
We assume weak turbulence in compressible medium follows similar scalings
as in incompressible one, which is a non-trivial
conjecture for low-$\beta$ medium.

\subsection{Incompressible limit: betatron acceleration}
In incompressible limit the slow modes become pseudo-Alfvenic perturbations
that move with the Alfven velocity. We should keep in mind
that for weak turbulence slow mode
perturbations have velocities substantially less than the Alfven velocity.
To calculate $D_p$ in weak turbulence,
we can still use the relation $\Delta p \sim (dp/dt)\Delta t 
                 \sim (pv_{l,slow}/l_{\|})(l_{\|}/v_{ptl})$ and 
$\Delta t \sim l_{\|}/v_{ptl}$ (see eq. (\ref{eq_Dp_main})).
However, anisotropy and scaling of $v_{l\|}$ are different.
Weak turbulence has an extreme anisotropy, $l_{\|}=$ constant, and $v_l \sim l^{1/2}$ (Galtier et al. 2000).
We have
\begin{eqnarray}
   D_p^{incom,weak} \sim (\Delta p)^2 /\Delta t 
   \nonumber \\
   \sim p^2 v_{L,slow}^2 (l_{mfp}/L) /(6v_{ptl} L_{\|})   
   \nonumber \\
   \sim \frac{ p^2 v_A }{ L } \left( \frac{ v_{L,slow} }{ v_A } \right)^2
        \frac{ l_{mfp}v_A }{ L_{\|} v_{ptl} },
\end{eqnarray}
where we used $l_{\|} \sim L_{\|}$.
This is smaller than $D_p^{incomp,strong}$ in eq. (\ref{Dpl}) by
a factor of $\sim l_{mfp}/L$ if $L_{\|} \sim L$.

\subsection{Slow modes: fast diffusion limit in compressible fluid}
The acceleration processes by weak and strong turbulence in the fast
diffusion limit are very similar. As in \S4.1  
consider $D_{\|}\gg Lc_s$, where $c_s$ is the propagation speed
of the slow wave and
depends on the plasma $\beta$: $c_s \sim v_A$ for
high-$\beta$ and $c_s \sim a$ (=sound speed) for low-$\beta$,
where we ignored the angle dependence for simplicity.
Inserting
\begin{equation}
  \Delta p \sim (pv_{l,slow}/l_{\|})(l_{\|}^2/D_{\|}) 
           \sim pv_{l,slow}l_{\|}/D_{\|}
\end{equation}
and
\begin{equation}
  \Delta t \sim l_{\|}^2/D_{\|}
\end{equation}
into $D_p \sim (\Delta p)^2/\Delta t$, we get
\begin{equation}
   D_{p,l} 
   \sim p^2 v_{l,slow}^2/D_{\|},       \label{eq_weak_fd}
\end{equation}
which is largest on the outer scale.

When we take into account multiple re-entry of the particle to
the same eddy, we need to multiply $D_p$ above by 
   $N_{w}\sim \min\{\sqrt{ D_{\|}\tau_{l,rand}}/l, M_{l,slow}\}$, where
   $\tau_{l,rand}$ is the eddy randomization 
   time of the slow modes on scale $l$.
   In weak turbulence, $\tau_{l,rand}\sim \chi^{-2}t_{l,wave}
   \sim t_{l,eddy}^2/t_{l,wave} \propto l$, 
   where $\chi = b_ll_{\|}/B_0l
   \sim t_{l,wave}/t_{l,eddy}$ (GS95) and we used 
   $t_{l,eddy}\sim l/v_l \sim l^{1/2}$ and 
   $t_{l,wave}=l_{\|}/c_s=$ constant.
Here $b_l$ is strength of the fluctuating field on scale $l$.
   Therefore the first term in the parenthesis is proportional to
   $\sim l^{-1/2}$. On the other hand, the second term,
   witch describes the characteristic scale of field line divergence,
   should decrease when $l$ decreases. The reason is as follows:
  the separation of field lines will show a random-walk behavior
  on scales larger than $L$.
  For a single eddy on the outer scale, separation of two adjacent field
  lines is
  $\sim L_{\|}b/B_0$, which will be the average path of the random walk.
  To reach the distance of $L$, we need 
  $\sim (LB_0/L_{\|}b)^2\sim \chi_L^{-2}$ steps,
  where
  $\chi_L$ is $\chi$ evaluated on the outer scale.
  Therefore, the second term in the parenthesis, $M_{l,slow}$, evaluated
  on the outer scale is $\sim \chi_L^{-2}$.  
  On the other hand, weak turbulence will show transition to
  strong turbulence deep down in the inertial range.
  We know that the factor $M_{l,slow}$ is of order unity when turbulence
  is strong. To summarize, the factor $M_{l,slow}$ is very large on the
  outer scale and order unity deep down in the inertial range.
  The most logical conclusion from this is that the factor $M_{l,slow}$
  decreases as $l$ decreases.
  As the result, the product of 
  the momentum diffusion coefficient in eq. (\ref{eq_weak_fd})
  and the factor $N_{w}$ decreases as $l$ decreases.

  Therefore, the efficiency of particle acceleration is largest
  on the outer scale. On the outer scale, the factor $N$ is
\begin{eqnarray}
   N_{w} \sim \min\{ \sqrt{D_{\|}\chi_L^{-2} t_{L,wave}}/L, \chi_L^{-2} \}
     \nonumber \\
     \sim \min\{ \chi_L^{-1}\sqrt{t_{L,wave}/t_{L,diff}}, \chi_L^{-2} \}
\end{eqnarray}
Note that $\chi_L \ll 1$ in weak turbulence.
The first term is smaller when $t_{L,diff} > \chi^2 t_{L,wave}$.
The net diffusion coefficient is
\begin{equation}
      D_{p}  
   \sim p^2 (v_{L,slow}^2/D_{\|}) N_{w}, 
\label{62} 
\end{equation}
where the factor $N_{w}\gg 1$. 

All in all, in the fast diffusion limit the acceleration efficiency is
increased by the factor of $N_w$ which is larger than the factor $N$ for
the strong turbulence. However, while for strong turbulence the injection
velocity may be of the order of Alfven one, for weak turbulence the
injection velocity should be smaller than $v_A$.

\subsection{Slow modes: slow diffusion limit in compressible fluid}
In this subsection, we consider $D_{\|}\ll L c_s$.
As we discussed in earlier sections, as long as
$D_{\|} \ll l_{\|}c_s$,
random walk-like behavior is suppressed.
The random walk-like behavior is fully recovered only for the eddies
that satisfy $D_{\|} \sim l_{\|}c_s$.
For slow and fast modes in strong turbulence, it is possible to
find such eddies.
However, since $l_{\|}$ is constant in weak turbulence,
it is {\it not} possible to find such a scale that satisfy
$D_{\|} \sim l_{\|}c_s$.
Therefore, acceleration of particles by large-scale motions is
inefficient in weak turbulence when diffusion is slow.

As we mentioned earlier, down the inertial range weak turbulence undergoes
a transition to the strong one as $k_{\bot}$ increases. The acceleration
by such turbulent motions is given by eq. (\ref{eq_qtd}) for low-$\beta$ plasma
and eq. (\ref{41}) for high-$\beta$ plasma where the velocity
of slow modes at the transition to the regime of strong turbulence should
be used.   

\subsection{Fast modes: slow and fast diffusion limits}

Acceleration that is induced by fast modes in weak turbulence
is analogous to the acceleration by fast modes in strong turbulence.
Therefore the momentum diffusion coefficient in {\it
fast diffusion limit} is given
again by eq. (\ref{eq_fastm_fastd}), but instead of multiplying it
by the factor $N$, as in the case of strong turbulence we should
multiply it by a larger factor $N_w$ to account for a higher probability
of a particle to return back to the compressive eddy. For slow diffusion
limit, the contribution by large scales eddies is given by eq. (\ref{eq_ptuskin_re2}) and therefore 
sub-dominant. This is due to the fact that for weak turbulence the turbulent diffusion is slow. 
The actual contribution to acceleration
in the slow diffusion limit is made by eddies with size $l\sim D_{\|}/c_f$. 
The amplitude of the fast modes depends on the properties of driving. 

\section{Acceleration by pitch-angle scattering}

Here we consider acceleration by pitch-angle scattering.
We do not need to know detailed physics of pitch-angle scattering.
However, we assume that characteristic propagation speed 
of the scattering agent is $\sim v_A$.
When the other velocities are involved, appropriate correction is needed.
Pitch-angle scattering is possible by both compressible and incompressible
motions. In most cases, this is the scattering that determines the 
particle mean free path and the $D_{\|}$.

When a particle interacts resonantly with small-scale waves, both the pitch-angle, the angle between momentum 
${\bf p}$ and ${\bf B}_0$,
 and the particle momentum change (Jokipii 1966;
Skilling 1975; Schlickeiser \& Achatz, 1993; 
see also Shalchi et al. 2004).
This pitch-angle scattering results in a diffusion process in momentum 
space and the 
particle distribution function $f$ follows
\begin{eqnarray}
 \frac{ \partial f }{ \partial t } = 
  \frac{ \partial }{ \partial \mu } \left( D_{\mu \mu} 
                   \frac{ \partial f }{ \partial \mu } \right)
 +\frac{ \partial }{ \partial \mu } \left( D_{\mu p} 
                   \frac{ \partial f }{ \partial p } \right) \nonumber \\
 + \frac{  1 }{ p^2 }
   \frac{ \partial }{ \partial p } \left[ p^2 
      \left(  D_{p\mu} \frac{ \partial f }{ \partial \mu } 
            + D_{pp}   \frac{ \partial f }{ \partial p } \right) \right]
\end{eqnarray}
(Schlickeiser \& Achatz 1993), where $\mu=\cos\theta$ and $\theta$ is the pitch-angle. 

The standard assumption for the studies of the acceleration by large
scale compressions (see P88, C03) is that the diffusivity arises from
the the pitch-angle scattering, while the $D_{pp}$ arises from the large-scale 
compressions.
However, it is easy to see that this is not true.
Indeed, the pitch-angle scattering frequency is
\begin{equation}
 \nu= 2 D_{\mu \mu}/(1-\mu^2).
\end{equation}

The $\mu$-averaged momentum scattering coefficient $D_p$ is
\begin{equation}
   D_p^{pitch} \sim \int_{-1}^{1} d\mu D_{pp}
\end{equation}
for isotropic particle distribution.
In general, the  momentum diffusion coefficient $D_{pp}$ is 
$\sim p^2(v_A/v_{ptl})^2$ times the 
pitch-angle diffusion coefficient $D_{\mu \mu}$:
$D_{pp} \sim p^2(v_A/v_{ptl})^2D_{\mu \mu}$
(Skilling 1975; Chandran 2000; YL02).
In fact, there is a simple interpretation for this: during each 
back-scattering, the fractional momentum change $\Delta p/p$ 
is $\sim v_A/v_{ptl}$.
Therefore, roughly speaking,
\begin{equation}
   D_p^{pitch} \sim p^2 (v_A/v_{ptl})^2 \int_{-1}^{1} d\mu D_{\mu \mu}
       \sim p^2 (v_A/v_{ptl})^2 \nu,   \label{eq_d_pitch}
\end{equation}
where we assumed $\int d\mu D_{\mu \mu} 
        \sim \int d\mu D_{\mu \mu}/(1-\mu^2)$.
More precisely speaking, 
\begin{equation}
 D_p^{pitch} = 4 p^2 \left( \frac{ v_A }{ v_{ptl} } \right)^2
     \left< \frac{ 1-\mu^2 }{ 2 } \frac{ \nu_+ \nu_- }{ \nu_+ + \nu_- }
     \right>,
\end{equation}
where the angled bracket $\langle ... \rangle$ indicates an average over
the pitch-angle cosine $\mu$ and $\nu_+$ and $\nu_-$ are
the pitch-angle scattering rates by waves moving parallel and anti-parallel
to ${\bf B}_0$, respectively (Skilling 1975). 
This equation reduces to eq. (\ref{eq_d_pitch}) when
$\nu_+=\nu_-$, which means that parallel and anti-parallel waves have
equal power. In other words, eq. (\ref{eq_d_pitch}) is
valid when
turbulence is balanced.
Note that $D_p$ is smaller than that in eq. (\ref{eq_d_pitch})
when turbulence is imbalanced.
Noting $\nu \sim v_{ptl}^2/D_{\|} \sim v_{ptl}/l_{mfp}$, 
we can rewrite eq. (\ref{eq_d_pitch}) as
\begin{eqnarray}
   D_p^{pitch} \sim p^2 (v_{A}/v_{ptl})^2 \nu   \nonumber \\
   \sim p^2 (v_{A}/v_{ptl})^2 v_{ptl}/l_{mfp}    
   \sim p^2 v_A^2 / (v_{ptl}l_{mfp})  \nonumber \\
   \sim \frac{ p^2 v_A }{ L } 
        \frac{ Lv_A }{ l_{mfp}v_{ptl} }. \label{eq_pitch}
\end{eqnarray}

Alfven, slow and fast modes contribute to pitch angle scattering. 
However, due to inefficiency of scattering by Alfven and slow modes
the corresponding $l_{mfp}$ is very large. Therefore the fast modes
dominate pitch-angle scattering.
The pitch-angle scattering is comparable to slow or fast modes
acceleration by large scale compressions
in fast diffusion limit and dominates them
in slow diffusion limit in the absence of shock compression.
However, in case particles are scattered by 
a static magnetic field, e.g. magnetic mirrors created by molecular clouds
(see Chandran 2001), the pitch-angle scattering 
will not accelerate particles efficiently.

Pitch-angle scattering also occurs in weak turbulence.
To calculate $D_p$,
we can use eq. (\ref{eq_pitch}).
However, the pitch-angle scattering efficiency, hence $l_{mfp}$, 
is less 
%
than that in strong turbulence because
anisotropy is more pronounced and 
fluctuation of magnetic field is smaller in weak turbulence. 
%
Weak turbulence in fast diffusion limit is dominated by large-scale non-resonance acceleration
given by eq.~(\ref{62}).

Transit Time Damping (TTD) 
and gyroresonance both produce pitch angle scattering\footnote{Formally
speaking both TTD and gyroresonance are given by the same condition (see 
Melrose 1968):
$\omega-k_{\|}v_{\|}-n\Omega/\gamma$, where frequency $\omega$ and a wavevector
component parallel to magnetic field
$k_{\|}$ are the characteristics of the magnetic turbulence, while $v_{\|}$
and $\Omega$ are, respectively, the component of particle velocity parallel
to magnetic field and gyroresonance frequency of the particle. The peculiar
feature of the TTD is that it corresponds to $n=0$ and therefore does not
depend on particle frequency. This entails substantial differences. In terms
of calculations, for instance, while the contributions
from large-scale MHD motions are unphysical for gyroresonance, as those would
violate the adiabatic invariant conservation
(Chandran 2000; Yan \& Lazarian 2003;
Shalchi, Yan \& Lazarian 2005), they should be accounted for in 
TTD calculations (YL02).}.

Gyroresonance is based on the resonance between the frequency
of gyrorotations and the frequency at which the particle meets
bumps in the magnetic field. Therefore sufficiently small
scales matter. The TTD is essentially Cherenkov-type
interaction, which allows interactions of particles with large-scale 
turbulence.
As the result, TTD is important for the situations when turbulence is damped at
small scales. Note, that TTD, unlike gyroresonance,
 requires compressions and therefore is inefficient for
Alfvenic perturbations.

At any rate,
we expect that the effect of TTD is very similar to that of
gyroresonance scattering\footnote{TTD, unlike gyroresonance, 
does not randomize the distribution of particles. Instead, it
increases the parallel particle velocity and does not change the
perpendicular velocity. However, this eventually should result
in instabilities that randomize particle velocities.}.   
For instance, 
$D_{pp}$ is  
$\sim p^2(v_A/v_{ptl})^2$ times
 the pitch-angle diffusion coefficient 
$D_{\mu \mu}$: $D_{pp}^{TTD}\sim p^2(v_A/v_{ptl})^2 D_{\mu \mu}^{TTD}$
(Yan \& Lazarian 2004).

A frequent argument against the TTD is that such mechanism accelerates
the particles along the magnetic field and therefore gets quickly saturated
as the fraction of modes available for acceleration decreases. Two
effects act to counteract this, however. First of all, magnetic field
wandering (see Lazarian \& Vishniac 1999) changes the direction of
magnetic field in respect of wavevectors of fast modes. Moreover, the
anisotropic distribution with particles moving mostly along magnetic field
lines is subjected to instabilities that will randomize particles.
We note that, while TTD may be also driven by large-scale compressions, 
it is different from
the large-scale stochastic mechanisms we deal for the most part of the paper.
It does require a rather coherent interaction with the Fourier component of
the random field. A particle surfs over the wave rather than bounces back
and forth within a compressed region.

\section{Comparison of Acceleration Mechanisms}

Above we discussed different mechanisms of particle acceleration
that make use of turbulence. These mechanisms can be roughly divided
into two groups: pitch-angle, which include gyroresonance as well
as TTD and non-resonant mechanisms that rely on large-scale compressions.
Dealing with the second group of mechanisms we obtained the 
acceleration rates for compressible (fast and slow) 
modes in both weak and strong turbulence,
considered limits of slow and fast particle diffusion. 
In addition, we showed that, the compressions of magnetic
field in incompressible fluid can still provide acceleration of particles
over their mean free path.

We summarize the results for strong MHD turbulence in Table 1.
When $v_A \sim v_{L,slow} \sim v_{L,fast}$, we have
\begin{equation}
   D_p^{incom} < D_p^{slow} \lesssim D_p^{fast} \lesssim D_p^{pitch}.
\end{equation}
Relative magnitude of $ D_p^{pitch}$ and $D_p^{supersonic}$
varies.
In slow diffusion limit (i.e.~$t_{L,diff} > t_{L,eddy}$), 
depending on the value of $Q_{TD}$,
$D_p^{slow}$ can be
larger than $D_p^{fast}$ or $D_p^{pitch}$.
However, since the factor $Q_{TD}$
is no larger than a logarithmic factor, we do not consider
the possibility seriously.
Noting
\begin{equation}
  \left( \frac{ Lv_A }{ l_{mfp} v_{ptl} } \right)
 \sim \left( \frac{ t_{L,diff} }{ t_{L,eddy} } \right),
\end{equation}
we find that
the momentum diffusion coefficient
is no less than 
\begin{equation}
   D_p \sim p^2v_A/L
\end{equation}
for slow diffusion limit.
In this limit, $D_p^{pitch}$ is larger than $D_p^{slow}$ or $D_p^{fast}$.
On the other hand, in fast diffusion limit,
$D_p^{slow}$, $D_p^{fast}$, and $D_p^{pitch}$ are all similar and
they can be less than $p^2v_A/L$.

In slow diffusion limit, when we ignore the $Q_{TD}$ factor, slow
modes are less efficient than fast modes.
This is because slow modes have longer wave periods 
(see Appendix B).
As we mentioned earlier, 
$D_p \sim p^2 (\Delta p)^2 / \Delta t$.
Let us consider $\beta \sim 1$ case.
For slow modes 
$\Delta p \sim p (\nabla \cdot {\bf v}_l) \Delta t 
          \sim p (v_l/l_{\|})(l_{\|}/v_A) \sim p v_{l,slow}/v_A$. 
For fast modes 
$\Delta p \sim p (\nabla \cdot {\bf v}_l) \Delta t 
          \sim p (v_l/l)(l/v_A) \sim p v_{l,fast}/v_A$,
where we used $c_f \sim v_A$.
Therefore, there is no big difference in $(\Delta p)_{slow}$ and
$(\Delta p)_{fast}$.
However, they have different $\Delta t$.
For slow modes, we have $\Delta t \sim l_{\|}/v_A$ (GS95).
For fast modes, we have $\Delta t \sim l/v_A$ because fast modes
are wave-like.
Since $l_{\|} \gg l$ on small scales, 
fast modes have much shorter $\Delta t$ on small scales.
Therefore, fast modes are more efficient.

Since pitch-angle scattering is comparable to or
more efficient than slow or fast
modes, we estimate the value of $D_p^{pitch}$ for various astrophysical
objects. We first rewrite the momentum diffusion coefficient by
pitch-angle scattering as follows:
\begin{equation}
   D_p^{pitch} \sim \frac{ p^2 v_A }{ L } 
                    \left( \frac{ t_{L,diff} }{ t_{L,wave} } \right),
\end{equation}
where $t_{L,wave} \sim L/v_A$.
{}From the relation $D_p \sim (\Delta p)^2/\Delta t \sim p^2/\tau_{acc}$,
we can define the acceleration time as follows:
\begin{equation}
   \tau_{acc} \equiv \frac{ p^2 }{ D_p }
\end{equation}
which is the timescale for doubling the momentum.
The results are given in Table 2.

\section{Discussion} 

Our calculations in the paper above reveal that in fast diffusion limit,
namely, when a particle diffuses away from a compression on the time 
scale less than the relevant crossing time, weak and slow turbulence
accelerates particles in very similar way. The most notable
difference between the two types of turbulence is that the particle
is more likely to return back to the compression created by
weak turbulence. At the same time, in the
slow diffusion limit the differences in acceleration by weak and strong
turbulence are substantial. They stem from both the difference in
turbulent scaling and from the differences in turbulence diffusivity.

Our study allows to answer a number of important questions, e.g. we can answer
the question we posed at the beginning of our paper. For instance, similar
to the case of scattering discussed in YL02, fast modes should dominate 
particle acceleration (cf. C03). We find that in many cases small scale
turbulent motions (comparable with the particle gyroradius)
that scatter particles also accelerate the particles more efficiently than
the turbulent motions at large scales. TTD which is also efficient for 
acceleration of particles but utilizes both small scale and large scale
compressions does not explicitly rely on particle randomization through
scattering. Unlike the TTD, the stochastic mechanisms that deal with large
scale motions do rely on scattering. Therefore, in many instances it is
enough to limit the studies of MHD turbulent acceleration to gyroresonance
and TTD (see Petrosian, Yan \& Lazarian 2005). The particular cases when 
alternative mechanisms of stochastic acceleration are important will be 
discussed elsewhere.

Similarly to the study in Yan \& Lazarian (2004) we assumed above that the
turbulent energy is being injected at the large scales and it cascades to
small scales. One of us (AL) studies currently a possibility of an additional
energy injection that happens at small scales and the effects of this on
cosmic ray. For instance, streaming
instability (see Cesarsky 1980) may be a source of injection of energy in
the form of waves (see YL02, Farmer \& Goldreich 2004, Yan \& Lazarian 2004).
In this case, Alfvenic motions would be more efficient in scattering compared
to the predictions in YL02. Moreover, the gyroresonance acceleration by such
motions will be also efficient.

For the scattering calculations in Yan \& Lazarian (2004) the plasma damping
effects were systematically taken into account and proven to be very 
important. In our present paper the damping of turbulence is only implicitly
present through our defining of the scales where turbulence is present and
through the mean free path of the particle $l_{mfp}$. 
For many of the discussed mechanisms the damping is unimportant as they are 
based on the interactions with large-scale fluctuations. However, for the 
gyroresonance the damping is important and the results from Yan \& Lazarian
(2004) can be applied directly.

The results that we obtained in the paper  
are order of magnitude only. Although calculations for 
gyroresonance and TTD acceleration
using quasi-linear theory  are available (see Yan \& Lazarian
2004) we did not use more precise formulae.
 The merit of our work is that it provides a rough comparison between different mechanisms of
stochastic acceleration, allows us
to define the dominant processes of stochastic acceleration, and
determines the necessary conditions for the different mechanisms to operate.
We believe our rough treatment may be justified and 
considered 
 adequate for many purposes
in view of various uncertainties of the model of turbulence. For instance,
an uncertainty arises from  determining
the amplitudes of turbulent perturbations.
Indeed, the acceleration is caused by slow and
fast modes, while usually only total velocity dispersion is available
through observations (see Lazarian \& Pogosyan 2000, 2004 and references
therein).
Yet extra uncertainties stem from
the uncertainty of the 
the injection scale and the scale of transition from weak to strong turbulence,
if this transition takes place. 

We hope that some of the uncertainties can be clarified. 
For instance, potentially compressible and incompressible
motions can be separated by statistical analysis of spectral line
variations (Lazarian \& Esquivel 2003,  Esquivel \& Lazarian 2005). Observational studies of
magnetic field intensities and further progress in theory of MHD
turbulence should make the transition from weak to strong turbulence
more certain.  Therefore a more precise treatment
of the acceleration processes would be welcome in future.
%

%
In the paper we discussed the acceleration by Alfven,
slow and fast
MHD modes. The possibility of considering
these MHD modes separately from the rest of the MHD cascade
stems from the scaling formulae in CL03
that shows that the energy exchange between different MHD
modes drops along the cascade. 

Our results show that some issues in the earlier studies of
turbulent acceleration require
revisions. For instance, the non-resonant acceleration in the slow
diffusion limit is in general more efficient than in P88. In
addition, fast modes are usually more important for acceleration
than slow modes (cf. C03). Moreover, it is not good to disregard
resonant acceleration while considering the non-resonant one. The
scattering that is required by the non-resonant acceleration provides
the acceleration rates that are usually comparable or larger than
those arising
from the non-resonant processes.
%

We believe that the importance of the stochastic acceleration is underestimated 
in the literature. It looks that this sort of  acceleration is important at least for
Solar flares, clusters of galaxies and gamma-ray bursts 
(see Petrosian 2001; Lazarian et al. 2002; Brunetti et al. 2004; Petrosian,
 Yan \& Lazarian 2005; Brunetti \& Blasi 2005). In addition, stochastic acceleration can also be important for
acceleration of the particles within the turbulent reconnection regions (see
Lazarian 2005).

We also mention that the particle acceleration mechanisms discussed
above are
applicable not only to energetic particles, but also to charged dust grains.
Resonant scattering and TTD were applied to charged dust grains in 
Yan \& Lazarian (2003). Because of the relatively small velocities of
dust grains, the factor $(v_A/v_{ptl})^2$ is usually larger than 1. Therefore
the gyroresonance and TTD acceleration are more efficient than scattering and
randomization of grains, and the resonant processes dominate the MHD
acceleration of dust grains. In many cases, due to inefficient magnetic scattering, the mean free path of the grain is determined by gaseous or plasma 
damping. Such a motion is mostly ballistic with the adiabatic invariant of
the grain conserved. As a result, the large scale compressions change
grain velocities by a factor of order unity.

All the results above assume turbulence where random motions are homogeneously
distributed in space. Turbulence intermittency should affect the distribution
function of the accelerated particles. This issue, however, is beyond 
the scope of the present paper.

\section{Summary}

We have studied particle acceleration by weak and strong MHD turbulence. We considered both case when the particle diffusion is slow and fast compared to the
life-time of a compressible perturbation.

1. Efficiency of particle acceleration depends upon turbulence being
strong or weak at  the injection scale

2. Pitch-angle scattering in most cases dominates cosmic ray acceleration 
when particle diffusion is slow.

3. Pitch-angle scattering, slow modes, and fast modes
give similar acceleration efficiency when particle diffusion is fast.

4. Turbulent diffusion arising from hydromagnetic motions
allows more efficient non-resonant acceleration when particle
diffusion is slow.

5. For strong turbulence, fast modes dominate the
 non-resonant acceleration. 

6. High compressions arising in
 super-sonic MHD turbulence increase efficiency of non-resonant particles 
   acceleration.

7. Incompressible MHD turbulence can provide non-resonant acceleration
of particles
over mean free paths.

\begin{acknowledgments}  
We thank Huirong Yan and Andrey Beresnyak for useful
suggestions. 
We are grateful to Ben Chandran for elucidating discussions
of the issue of the parallel momentum loss
and for John Mathis for reading the manuscript. The research was supported by 
NSF grant ATM 0312282 and the NSF 
Center
for Magnetic Self Organization in Laboratory and Astrophysical Plasmas.
JC's study was financially supported by research fund of
Chungnam National University in 2005.
\end{acknowledgments}


\begin{deluxetable}{llll}  
\tabletypesize{\footnotesize}
\tablecaption{Summary of $D_p$}
\tablewidth{0pt}
\tablehead{
\colhead{$D_p$} & \colhead{Estimates}   
& \colhead{Properties of fluid}  & \colhead{Eq.}
}
\startdata
  $D_p^{incomp}$ & 
     $\sim \frac{ p^2 v_A }{ L } \left( \frac{ v_{L,slow} }{ v_A }
                                  \right)^2
       \left( \frac{ v_{A} }{ v_{ptl} } \right)$    &       
     incompressible fluid   &
      (\ref{Dpl})
 \\
 \\
   $D_p^{slow}$
      &  $\sim \frac{ p^2 v_A }{ L } \left( \frac{ v_{L,slow} }{ v_A }
                                  \right)^2
            \left( \frac{ Lv_A }{ l_{mfp}v_{ptl} } \right)    $    &
       slow modes, low-$\beta$ \tablenotemark{a)}~~,
  $t_{L,diff} \leq t_{L,eddy}$\tablenotemark{b)} &
      (\ref{eq_sm_fd1})
 \\
    &  $\sim \frac{ p^2 v_A }{ L } \left( \frac{ v_{L,slow} }{ v_A }
                         \right)^2 Q_{TD}$ \tablenotemark{c)}     &
       slow modes,  low-$\beta$,
       $t_{L,diff} \geq t_{L,eddy}$    &
       (\ref{eq_qtd})
 \\
      &  $\sim \frac{ p^2 v_A }{ L } \left( \frac{ v_{L,slow} }{ v_A }
                                  \right)^2
      \left( \frac{ Lv_A }{ l_{mfp}v_{ptl} } \right)  \beta^{-2}  $    &
       slow modes, high-$\beta$,
      $t_{L,diff} \leq t_{L,eddy}$    &
      (\ref{eq_ptuskin_fast2_hb})
 \\
    &  $\sim \frac{ p^2 v_A }{ L } \left( \frac{ v_{L,slow} }{ v_A }
                                  \right)^2  \beta^{-2} Q_{TD}$    &
       slow modes,  high-$\beta$,
       $t_{L,diff} \geq t_{L,eddy}$  
    & (\ref{eq_d_p_lc_hb})
 \\
 \\
    $ D_p^{fast}$ &   
    $\sim \frac{ p^2 v_A }{ L } \left( \frac{ v_{L,fast} }{ v_A }
                               \right)^2 
                                  \left( \frac{ Lv_A }{ l_{mfp}v_{ptl} }
                                  \right)^{} $   &
          fast modes, all-$\beta$, 
    $t_{L,diff} \leq t_{L,wave}$ &
    (\ref{eq_fastm_fastd})
 \\
 \\
    &  $\sim \frac{ p^2 v_A }{ L } \left( \frac{ v_{L,fast} }{ v_A }
                               \right)^2 
                                  \left( \frac{ Lv_A }{ l_{mfp}v_{ptl} }
                                  \right)^{1-2m} $   &
          fast modes, low-$\beta$,
   $t_{L,diff} \geq t_{L,wave}$ &
   (\ref{eq_fastm_slowd_lb})
 \\
 \\ 
    &  $\sim \frac{ p^2 v_A }{ L } \left( 
       \frac{ v_{L,fast} }{ v_A } 
                                  \right)^2
                                  \left( \frac{ Lv_A }{ l_{mfp}v_{ptl} }
                                  \right)^{1-2m}  
                                 \beta^{-m}
                 $   &
       fast modes, high-$\beta$, 
   $t_{L,diff} \geq t_{L,wave}$  &
   (\ref{eq_fastm_slowd_hb})
 \\  
 \\
 $ D_p^{pitch}$ & 
      $\sim \frac{ p^2 v_A }{ L } 
      \left(  \frac{ Lv_A }{ l_{mfp}v_{ptl} }  \right)  $   &
       pitch-angle scattering &
      (\ref{eq_pitch})
 \\
 \\
 $ D_p^{supersonic}$ &
       $\sim \frac{ p^2 v_A }{ L }
         \left(   \frac{ v_L }{ v_A }  \right)
            M_s^{2\gamma} $    &
       super-sonic fluids \tablenotemark{d)}  &
       (\ref{eq_supersonic})
 \\
\enddata
\tablenotetext{a)}{The plasma $\beta$ is defined by the ratio of the
   gas pressure and the magnetic pressure: 
   $\beta = P_g/P_{mag}$ ~$(= a^2/v_A^2$ for isothermal gas), 
   where $a$ is the sound speed and $v_A$ the Alfven speed.
   The value of $m$ is related to the scaling of
   fast mode: $v_{l,fast}\sim v_{L,fast} (l/L)^m$.
}
\tablenotetext{b)}{Eddy turnover time at the
  outer scale $L$ is $t_{L,eddy}\sim L/v_A$; diffusion time
  at the scale is $t_{L,diff}\sim L^2/D_{\|} \sim L^2/l_{mfp}v_{ptl}$;
  wave period at the outer scale is $t_{L,diff}\sim L/c_f$,
  where $c_f$ is the propagation speed of fast modes.
  Note that $t_{L,diff}/t_{L,eddy} \sim Lv_A/D_{\|} \sim 
                   Lv_A/l_{mfp}v_{ptl}$;
                   $t_{L,diff}/t_{L,wave} \sim Lc_f/D_{\|} \sim 
                   Lc_f/l_{mfp}v_{ptl}$.
      Here $l_{mfp}$ is the mean free path of particle back-scattering
      by pitch-angle scattering.
}
\tablenotetext{c)}{$Q_{TD} \sim 1$ when turbulent diffusion is slow 
   and $Q_{TD}\sim \ln(Lv_A/D_{\|})$ when it is fast.
}
\tablenotetext{d)}{The parameter $\gamma$ is related to 
  the compression rate of the fluid: $M_s^{\gamma}$ is the ratio of
  the density in a typical compressed region and average density,
  where$M_s$ is the sonic Mach number. 
  We expect  $\gamma < 2$ in MHD supersonic turbulence, 
  while $\gamma \sim 2$ in hydrodynamic counterpart.
}
 \label{table_1}
\end{deluxetable}


\begin{deluxetable}{lll}  
\tabletypesize{\scriptsize}
\tablecaption{Estimates of $D_p$}
\tablewidth{0pt}
\tablehead{
 \colhead{Astrophysical fluids} & 
 \colhead{Estimates of $D_p^{pitch}$ \tablenotemark{a)}}   
 & \colhead{Estimates of $\tau_{acc}$ \tablenotemark{b)}}  
}
\startdata
   Solar flares   &        
    $ (2 \mbox{~s$^{-1}$}) p^2    
    \left( \frac{ v_A }{ 10^4 km/sec } \right)
    \left( \frac{ 5000 km }{ L }\right)
    \left( \frac{ t_{L,diff} }{ t_{L,wave} } \right) $  &
   $ 0.5 \mbox{~s}                                  
    \left( \frac{ 10^4 km/sec }{ v_A } \right)
    \left( \frac{ L }{ 5000 km } \right)
    \left( \frac{ t_{L,wave} }{ t_{L,diff} } \right) $
\\
\\
   ICM &
  $ (3\times 10^{-8}  \mbox{~yrs$^{-1}$})p^2   
    \left( \frac{ v_A }{ 300 km/sec } \right)
    \left( \frac{ 10 kpc }{ L }\right) 
    \left( \frac{ t_{L,diff} }{ t_{L,wave} } \right) $ &
  $ 3 \times 10^7 \mbox{~yrs}                                  
    \left( \frac{ 300 km/sec }{ v_A } \right)
    \left( \frac{ L }{ 10 kpc } \right)
    \left( \frac{ t_{L,wave} }{ t_{L,diff} } \right) $
\\
\\
   Gamma-ray bursts & 
  $ (30  \mbox{~s$^{-1}$})p^2   
    \left( \frac{ v_A }{ 3\times 10^5 km/sec } \right)
    \left( \frac{ 1000 km }{ L }\right)
    \left( \frac{ t_{L,diff} }{ t_{L,wave} } \right)  $   &
  $ 0.03 \mbox{~s}                                  
    \left( \frac{ 3\times 10^5 km/sec }{ v_A } \right)
    \left( \frac{ L }{ 1000 km } \right)
    \left( \frac{ t_{L,wave} }{ t_{L,diff} } \right)  $
\\
\\
   Galactic halo\tablenotemark{c)} &
  $  (10^{-6}  \mbox{~yrs$^{-1}$})p^2   
    \left( \frac{ v_A }{ 100 km/sec } \right)
    \left( \frac{ 100 pc }{ L }\right)
    \left( \frac{ t_{L,diff} }{ t_{L,wave} } \right)  $  & 
  $   10^6 \mbox{~yrs}                                  
    \left( \frac{ 100 km/sec }{ v_A } \right)
    \left( \frac{ L }{ 100 pc } \right)
    \left( \frac{ t_{L,wave} }{ t_{L,diff} } \right)  $
\\
\enddata
\tablenotetext{a)}{
    $ D_p^{pitch} \sim \frac{ p^2 v_A }{ L } 
                    \left( \frac{ t_{L,diff} }{ t_{L,wave} } \right) $
}
\tablenotetext{b)}{
    $ \tau_{acc} \equiv \frac{ p^2 }{ D_p } $
}
\tablenotetext{c)}{
    For Galactic halo, $l_{mfp} \sim 10^{20}$ cm (Yan \& Lazarian 2004)
    for particles with energy larger than $\sim 1 GeV$.
    The diffusion time is $\sim L^2/(l_{mfp}c)
          \sim (100pc)^2/(l_{mfp}c) \sim 1000$ yrs.
    The wave period is $\sim L/v_A \sim 100pc/100km/sec 
          \sim 10^6$ yrs.
    Cosmic rays in Galactic halo are in fast diffusion limit.
    We have $t_{L,wave} / t_{L,diff} \sim 10^3$
    and $\tau_{acc} \sim 10^9$ yrs.
}
\label{tabel_2}
\end{deluxetable}

\appendix

\section{Calculation of $D_p$} 

\subsection{$D_p$ from anisotropic slow modes}

 Chandran \& Maron (2004b) considered particle acceleration
by strong turbulence. They assumed that   
the eddy turnover time $t_{eddy}\sim L/v_L$) is less than the
 particle diffusion 
time $t_{diff}\sim L^2/D$, i.e. that the particles can diffuse out of an eddy
before it is randomized. The corresponding 
$\Delta t =  t_{diff}$ and
$\nabla \cdot {\bf v}_L \sim v_L/L$ and obtained 
(see \S\ref{sect_hydro_method})
\begin{equation}
   D_p \sim p^2 (v_L^2/L^2)(L^2/D) \sim p^2 v_L^2/D \sim p^2 v_A^2/D,
\end{equation}
where $v_L \sim v_A$ is assumed.
They observed that, when particles diffuse along
magnetic field lines, they can reenter the same outer scale eddy 
\begin{equation}
   N\sim min\{\sqrt{ D_{\|}\tau_{rand}}/L, M\}    \label{eq_N}
\end{equation}
times\footnote{
   The first term on the right is due to diffusion process.
   For a diffusion process, we can use the relation,
   $ d^2 \sim D_{\|} \Delta t$, where $d$ is the average 
   distance the particle travels
   through diffusion.
   During one eddy turnover time of the outer scale, $L/v_L$,
   the test particle can travel 
   $d \sim \sqrt{D_{\|}L/v_L}$.
   The number we are looking for is 
    $d/L\sim \sqrt{D_{\|}/Lv_L}\sim \sqrt{D_{\|}L/v_L}/L$.
}
before the outer scale is randomized.
Here
   $\tau_{rand}$ is the eddy randomization time on the outer scale
   of turbulence.
The factor $M$ is defined
    as $z_s/L$,
   where $z_s$ is the distance along magnetic field lines over which two
   adjacent magnetic field lines get separated by the distance $L$.
   The estimates for this factor range from 1 (Narayan \& Medvedev 2001)
   to 5 or 7 (Chandran \& Maron 2004b). 
   Therefore the value of $N$ is no more than a few in strong turbulence.
  We observe that intrinsic diffusion
  of particles due to scattering should decrease $M$, while if turbulence
   is weak, $M$ should increase.

On the other hand, when $t_{diff} > t_{eddy}$, 
particles are confined within the eddies until the eddies are
randomized.
C03 used 
$\nabla \cdot {\bf v}_L \sim v_L/L$ and 
$\Delta t \sim t_{eddy}$ ($\sim L/v_L$) for outer scale eddies
and obtained
\begin{equation}
     D_p \sim p^2(v_L^2/L^2)(L/v_L) \sim p^2 v_L/L  \sim p^2 v_A/L
\end{equation}
(see also \S\ref{sect_hydro_method}).
C03 claimed that the
small scale structure of slow mode turbulence is also
important when $d_{min}v_A \ll D_{\|} \ll Lv_A$, where $v_A$ is the
Alfven speed and
$d_{min}$ is either the parallel size of eddies at the dissipation scale
or the mean free path of the cosmic rays, 
whichever is larger.
Particle diffusion time is smaller than the eddy turnover time at the 
outer scale $L$ and larger than that at the scale $d_{min}$.
For eddies whose parallel size is larger than $D_{\|}/v_A$, 
particles are confined within the eddies until the eddies are
randomized. For such eddies, we note that
$\nabla \cdot {\bf v}_{l,slow}
\sim (v_{l,slow}/l) (\hat{\bf k} \cdot {\bf \xi}_s) 
\sim v_{l,slow}/l_{\|}$
(see Appendix B1 for details) and 
$\Delta t \sim l_{\|}/v_A$ (GS95).
Therefore (see eq. (\ref{eq_univ})), 
\begin{equation}
  D_{p,l}^{C03} \sim (\Delta p)^2/\Delta t 
   \sim (p^2v_{l,slow}^2/v_Al_{\|})
   \sim (p^2 v_{L,slow}^2/v_A L),
\label{Dp_c03}
\end{equation}
where we used $v_{l\|} \sim v_{L,slow} (l/L)^{1/3}$ and $l_{\|} \sim L^{1/3}l^{2/3}$.
Each eddy whose parallel size is between $D_{\|}/v_A$ and $L$ 
makes an equal contribution to $D_p$.
Since there are roughly 
$\int dl/l \sim \ln\left( { Lv_A }/{ D_{\|} } \right)$ eddies in
the parallel size range,
the sum of all contributions from such eddies is
\begin{equation}
   D_p^{C03} \sim \frac{ p^2 v_A }{ 6L }  \left(\frac{v_{L,slow}}{v_A}\right)^2
    \ln\left( \frac{ Lv_A }{ D_{\|} } \right),
\label{main}
\end{equation}
which for $v_{L,slow}\sim v_A$ is equal to eq. (30) in 
C03. Here we ignored constants of order unity.
Note that $D_p^{C03}$ given above does not go to zero
when $D$ goes to zero. 
Therefore, this result is in disagreement with
Ptuskin's result in eq. (\ref{eq_ptuskin}).

\subsection{Acceleration by electric field arising from slow modes
   in incompressible limit}
In \S\ref{sect_2}, we showed that there is an exact 
cancellation between energy gain/loss by electric field in the
perpendicular directions and adiabatic  gain/loss in the parallel
directions {\it when particle distribution is isotropic}.
Suppose that, somehow, particle distribution is anisotropic.
In this case, we can use eq. (\ref{eq_dpdt}) for $dp/dt$ and
$l_{\|}/v_A$ for $\Delta t$. 
Therefore, the resulting $D_p$ will be very similar to
those in \S\ref{sect_slow}, which means that particle acceleration
will be efficient {\it if} particle distribution is anisotropic.

\section{Scaling of slow and fast modes}
\subsection{Calculation of $\nabla \cdot {\bf v}$}  
In Fourier space, $\nabla \cdot {\bf v}$ becomes 
$i{\bf k} \cdot {v}_{\bf k} \hat{\bf \xi}$, where $i^2=-1$, ${\bf k}$
is the wave-vector,
$ {v}_{\bf k}$ is the Fourier component of velocity, and 
$\hat{\bf \xi}$ is
the unit vector along the direction of $\hat{\bf v}_{\bf k}$.
Cho \& Lazarian (CL03) showed that
\begin{eqnarray}
   \hat{\bf \xi}_s \propto 
     ( -1 + \alpha - \sqrt{D} )
            k_{\|} \hat{\bf k}_{\|} 
     + 
     ( 1+\alpha - \sqrt{D} ) k_{\perp} \hat{\bf k}_{\perp},
  \label{eq_xis_new2}
\\
   \hat{\bf \xi}_f \propto 
     ( -1 + \alpha + \sqrt{D} )
            k_{\|} \hat{\bf k}_{\|} 
     + 
     ( 1+\alpha + \sqrt{D} ) k_{\perp} \hat{\bf k}_{\perp},
   \label{eq_xif_new2}
\end{eqnarray}
where $\alpha = a^2/v_A^2 = \beta/2$ for isothermal gas, 
$D=(1+\alpha)^2-4{\alpha} \cos^2{\theta}$,
$\cos \theta = k_{\|}/k$, and $k_{\perp}$ is
wave number perpendicular to the mean field and $k_{\|}$
parallel to it.
Hatted vectors are unit vectors.
The slow basis $\hat{\bf \xi}_s$ lies between $\hat{\bf k}_{\|}$ and
$-\hat{\bf \theta}$.
The fast basis $\hat{\bf \xi}_f$ lies between $\hat{\bf k}_{\perp}$ and
$\hat{\bf k}$.
Here overall sign of $\hat{\bf \xi}_s$ and $\hat{\bf \xi}_f$ is not important.

When $\beta \rightarrow 0$,
equations (\ref{eq_xif_new2}) and (\ref{eq_xis_new2}) becomes
\begin{eqnarray}
   \hat{\bf \xi}_s \approx  \hat{\bf k}_{\|}
     -(\alpha \sin\theta \cos\theta)\hat{\bf k}_{\perp} 
   \rightarrow \hat{\bf k}_{\|}, \label{xis_lowbeta}
\\
   \hat{\bf \xi}_f \approx  (\alpha \sin\theta \cos\theta) \hat{\bf k}_{\|}
                         +\hat{\bf k}_{\perp}
   \rightarrow \hat{\bf k}_{\perp}.    \label{xif_lowbeta}     
\end{eqnarray}
In this limit, $\hat{\bf \xi}_s$ is mostly proportional to $\hat{\bf k}_{\|}$
and $\hat{\bf \xi}_f$ to $\hat{\bf k}_{\perp}$.
When $\alpha \rightarrow \infty$,
equations (\ref{eq_xif_new2}) and (\ref{eq_xis_new2}) becomes
\begin{eqnarray}
   \hat{\bf \xi}_s \approx  -\hat{\bf \theta}
             +(\sin\theta \cos\theta/\alpha)\hat{\bf k}, \label{xis_highbeta}
\\
   \hat{\bf \xi}_f \approx
                 (\sin\theta \cos\theta/\alpha) \hat{\bf \theta}
                                   +\hat{\bf k}.    \label{xif_highbeta}
\end{eqnarray}
When $\alpha = \infty$, slow modes are called {\it pseudo}-Alfvenic modes.

The quantity $\nabla \cdot {\bf v}_l$ becomes
$\sim (v_l/l) \langle \hat{\bf k} \cdot \hat{\bf \xi} \rangle$, where
$\langle ... \rangle$ denotes average over Fourier components
that have $k\sim 1/l$.
For slow modes in low-$\beta$ plasmas,
$\hat{\bf k} \cdot \hat{\bf \xi}_s \sim \hat{\bf k} \cdot \hat{\bf k}_{\|}
   \sim k_{\|}/k$ and, therefore,
\begin{equation}
   \nabla \cdot {\bf v}_{l,slow} \sim v_{l,slow}/l_{\|}.
   \label{eq_divslow_app}
\end{equation}
For slow modes in high-$\beta$ plasmas,
$\hat{\bf k} \cdot \hat{\bf \xi}_s \sim 
   \hat{\bf k} \cdot (\sin\theta \cos\theta /\alpha)\hat{\bf k}
   \sim (k_{\|}/k)/\alpha$ and, therefore,
\begin{equation}
   \nabla \cdot {\bf v}_{l,slow} \sim v_{l,slow}/(\beta l_{\|}).
\end{equation}
For fast modes
\begin{equation}
   \nabla \cdot {\bf v}_{l,fast} \sim v_{l,fast}/l
  \label{eq_divfast_app}
\end{equation}
for all values of $\beta$.
Since $l_{\|} \geq l$, the value in eq. (\ref{eq_divfast_app}) 
is larger than that in (\ref{eq_divslow_app}).
This explains why fast modes are more efficient in particle
acceleration.

\subsection{Estimates for density fluctuations}
In CL03, we provided estimates for density 
fluctuations in MHD turbulence.
In case there are only slow modes, the density fluctuations are
\begin{equation}
 (\delta \rho / \bar{\rho} )_{slow} 
    \sim v_{L,slow}/a 
    \label{eq_slow_den} 
\end{equation}
in low-$\beta$ plasmas
(CL03).
However, when Alfven modes are present, slow modes are
enslaved by Alfven modes (see discussions in Lithwick \& Goldreich 2001).
In this case, eq. (\ref{eq_slow_den}) is modified to
\begin{equation}
 (\delta \rho / \bar{\rho} ) \sim v_{L,slow}/v_A,
\end{equation}
which is $\sim O(1)$ when $v_{L,slow}\sim v_A$.
The contribution from fast modes is
$(\delta \rho / \bar{\rho} )_{fast} 
    \sim v_{L,fast}/v_A$.
Therefore, when $v_{L,fast} < v_{L,slow}$,
slow modes dominate density fluctuations in low-$\beta$ plasmas.
Although fast modes can compress more 
(c.f.~eq. (\ref{eq_divfast_app})), short time scale, or wave period, 
of fast modes makes them not so efficient.

In high-$\beta$ plasmas, fast modes dominate density fluctuations:
\begin{equation}
 (\delta \rho / \bar{\rho} ) \sim v_{L,fast}/a.
\end{equation}
When $v_{L,fast}\sim v_A$, we have 
$(\delta \rho / \bar{\rho} ) \sim 1/\sqrt{\beta}$.







\begin{thebibliography}{8.}
\addcontentsline{toc}{section}{References}

\bibitem{} Armstrong, J.W., Rickett, B.J., \& Spangler, S.R., 1995, 
ApJ, 443, 209
\bibitem{BykT82} Bykov, A. \& Toptygin, I. 1982, 
     J. of Geophysics-Zeitschrift fuer Geophysik, 50, 221
\bibitem{BreLC05} Beresnyak, A., Lazarian, A., \& Cho, J. 2005, ApJ, 624, L93
\bibitem{} Brunetti, G., Blasi, P., Cassano, R., \& Gabilci, S. 2004, MNRAS, 320, 365
\bibitem{} Brunetti, G. \& Blasi, P. 2005, MNRAS, in press 

\bibitem{Ces80} Cesarsky, C. 1980, ARA\&A, 18, 289
\bibitem{Cha00} Chandran, B. 2000, Phys. Rev. Lett., 85, 4656
\bibitem{Cha01} Chandran, B. 2001, Sp. Sci. Rev., 99, 271
\bibitem{Cha03} Chandran, B. 2003, ApJ, 599, 1426 (C03)
\bibitem{ChaM04a} Chandran, B. \& Maron, J. 2004a, ApJ, 602, 170
\bibitem{ChaM04b} Chandran, B. \& Maron, J. 2004b, ApJ, 603, 23 

\bibitem{ChoL02} Cho, J. \& Lazarian, A. 2002, Phys. Rev. Lett., 24, 5001
\bibitem{ChoL03} Cho, J. \& Lazarian, A. 2003, MNRAS, 345, 325 (CL03)
\bibitem{ChoL04} Cho, J. \& Lazarian, A. 2004, in Studying Turbulence
Using Numerical Simulations, Stanford, p. 75
\bibitem{ChoV00} Cho, J. \& Lazarian, A. 2005, Theo. Comp. Fluid Mech., 
                 19, 127
\bibitem{ChoLH03} Cho, J., Lazarian, A., Honein, A., Knaepen, B.,
                  Kassinos, S., \& Moin, P. 2003, ApJ, 589, L77
\bibitem{ChoV00} Cho, J., Lazarian, A., \& Vishniac, E. 2002, ApJ, 564, 291
\bibitem{ChoV00} Cho, J., Lazarian, A., \& Vishniac, E. 2003,
              in {\it Turbulence and magnetic fields in astrophysics}, 
                 Lect.~Notes~Phys. Vol.~614,
      eds. E. Falgarone \&  T. Passot (Berlin: Springer, 2003), p. 56 
\bibitem{ChoV00} Cho, J. \& Vishniac, E. 2000, ApJ, 539, 273

\bibitem{FarG04} Farmer, A. \& Goldreich, P. 2004, ApJ, 604, 671
\bibitem{Fer49} Fermi, E. 1949, Phys. Rev. 75, 1169

\bibitem{GalP00} Galtier, S., Nazarenko, S., Newell, A., \&
                 Pouquet, A. 2000, J. Plasma Phys., 63, 447 
\bibitem{GoldS95} Goldreich, P. \& Sridhar, S. 1995, ApJ, 438, 763 (GS95)

\bibitem{Hig84} Higdon, J. C. 1984, ApJ, 285, 109
\bibitem{Jok66} Jokipii, R. 1966, ApJ, 146, 480
\bibitem{Jok68} Jokipii, R. 1968, ApJ, 152, 997

\bibitem{KimR05} Kim, J. \& Ryu, D. 2005, ApJL, 630, L45
\bibitem{KulF71} Kulsrud, R. \& Ferrari, A. 1971, Astrophys. Sp. Sci. 
                 12, 302

\bibitem{} Lazarian, A. 2005, in Magnetic Fields in the Universe,
eds, E. Gouveia Dal Pino, G. Lugones, A. Lazarian, AIP 784, p. 42
\bibitem{LazE03} Lazarian, A., \& Esquivel, A. 2003, ApJ. Lett., 592, 37
\bibitem{} Lazarian, A., Petrosian, V., Yan, H. \& Cho, J. 2002, Beaming and Jets in Gamma
Ray Bursts, eds. Rachid Ouyed, Copenhangen, p. 45
\bibitem[]{2368} Lazarian, A., \& Pogosyan D. 2000, ApJ, 573, 720 
\bibitem[]{2369} Lazarian, A., \& Pogosyan D. 2004, ApJ, 616, 943
\bibitem{LazV99} Lazarian, A. \& Vishniac, E. 1999, ApJ, 517, 700
\bibitem{LitG01} Lithwick, Y. \& Goldreich, P. 2001, ApJ, 562, 279

\bibitem{} Melrose, D.B., 1968, ApSS, 2, 171

\bibitem[]{2374} Montgomery, D., Brown, M. R., \& Matthaeus, W. H. 1987, J. Geophys. Res. 92, 282
\bibitem[]{2375} Narayan, R. \& Medvedev, M. 2001, ApJ, 562, 129
\bibitem[]{2376} Padoan, P., Nordlund, A., \& Jones, B. 1997, MNRAS, 288, 145
\bibitem[]{2377} Padoan, P. \& Nordlund, A. 1999, ApJ, 526, 279
\bibitem[]{2378} Padoan, P. \& Nordlund, A. 2002, ApJ, 576, 870
\bibitem{} Petrosian, V. 2001, ApJ, 557, 560
\bibitem[]{2379} Petrosian, V., \& Liu, S. 2004, ApJ, 610, 550
\bibitem{} Petrosian, V., Yan, H. \& Lazarian, A. 2005, ApJ, submitted

\bibitem{Ptu88} Ptuskin, V. 1988, Soviet Astron. Lett. 14, 255 (P88)

\bibitem{SauPP02} Saur, J., Politano, H., Pouquet, A., \&
                  Matthaeus, W. 2002, A\&A, 386, 699
\bibitem{Sch89} Schlickeiser, R. 1989, ApJ, 336, 243
\bibitem{SchA93} Schlickeiser, R. \& Achatz, U. 1993, J. Plasma Phys., 49,
                63
\bibitem{SchM98} Schlickeiser, R. \& Miller, J. 1998, ApJ, 492, 352

\bibitem{ShaBM04} Shalchi, A., Bieber, J., Matthaeus, W., \& Qin, G.
                  2004, ApJ, 616, 617
\bibitem{}  Shalchi, A., Yan, H., \& Lazarian A. 2005, MNRAS, 356, 1064 
\bibitem{SheMM83} Shebalin, J. V., Matthaeus, W., \& Montgomery, D. 
                  1983, J. Plasma Phys., 29, 525
\bibitem{SheM88} Shebalin, J. V., \& Montgomery, D. 1988, J. Plasma Phys. 39, 339

\bibitem{Ski75} Skilling, F. 1975, MNRAS, 172, 557

\bibitem{YanL02} Yan, H. \& Lazarian, A. 2002, Phys. Rev. Lett., 89, 281102 (YL02)
\bibitem[]{2396} Yan, H. \& Lazarian, A. 2003, ApJL, 592, L37
\bibitem{YanL04} Yan, H. \& Lazarian, A. 2004, ApJ, 614, 757   

\bibitem{ZakS70} Zakharov, V.E. \& Sagdeev, A. 1970, Sov. Phys. Dokl.,
                 15, 439
\bibitem[]{2401} Zank, G.P. \& Matthaeus, W. H. 1992, J. Plasma Phys. 48,85

\end{thebibliography}
\end{document}